\documentclass[letterpaper,twocolumn,10pt,accepted=2023-08-16]{quantumarticle}
\pdfoutput=1
\usepackage[utf8]{inputenc}
\usepackage[english]{babel}
\usepackage[T1]{fontenc}
\usepackage{amsmath, amssymb}
\usepackage{hyperref}

\usepackage{tikz}
\usepackage{lipsum}

\usepackage{graphicx} 
\usepackage{subfigure}
\usepackage[normalem]{ulem}
\usepackage{longtable} 
\usepackage{bm} 

\usepackage{tikz}
\usetikzlibrary{decorations.pathreplacing, decorations.pathmorphing, arrows, positioning, calc, intersections, shapes, matrix, fit}

\def\>{\rangle}
\def\<{\langle}

\def\->{\rightarrow}
\def\=>{\implies}
\def\6{\partial}
\def\Tr{\text{Tr}}

\newcommand{\bra}[1]{\ensuremath{\left\langle{#1}\right\vert}}
\newcommand{\ket}[1]{\ensuremath{\left\vert{#1}\right\rangle}}

\renewcommand{\Re}{\operatorname{Re}}
\renewcommand{\Im}{\operatorname{Im}}
\newcommand{\ip}[2]{\langle #1 | #2 \rangle}

\def\be{\begin{equation}}
\def\ee{\end{equation}}
\newcommand{\beq}{\begin{equation}}
\newcommand{\eeq}{\end{equation}}
\newcommand{\bqa}{\begin{eqnarray}}
\newcommand{\eqa}{\end{eqnarray}}

\newcommand{\BQIC}{Berkeley Center for Quantum Information and Computation, Berkeley, California 94720 USA}
\newcommand{\DeptPhys}{Department of Physics, University of California, Berkeley, California 94720 USA}
\newcommand{\DeptChem}{Department of Chemistry, University of California, Berkeley, California 94720 USA}
\newcommand{\MCQE}{Center for Engineered Quantum Systems,  Macquarie University, 2109 NSW, Australia}
\newcommand{\Tech}{Department of Physics, Technion - Israel Institute of Technology, Haifa 32000 Israel}

\newcommand{\erf}[1]{Eq.~(\ref{#1})}
\newcommand{\erfs}[1]{Eqs.~(\ref{#1})}

\begin{document}

\title{A Multi-Qubit Quantum Gate Using the Zeno Effect}

\author{Philippe Lewalle}
\email{plewalle@berkeley.edu}
\affiliation{\BQIC}
\affiliation{\DeptChem}
\author{Leigh S. Martin}
\affiliation{\BQIC}
\affiliation{\DeptPhys}
\thanks{PL and LSM contributed equally to this manuscript.}
\author{Emmanuel Flurin}
\affiliation{\BQIC}
\affiliation{\DeptPhys}
\author{Song Zhang}
\affiliation{\DeptChem}
\author{Eliya Blumenthal}
\affiliation{\Tech}
\author{Shay Hacohen-Gourgy}
\affiliation{\Tech}
\author{Daniel Burgarth}
\affiliation{\MCQE}
\author{K. Birgitta Whaley}
\affiliation{\BQIC} 
\affiliation{\DeptChem}

\date{November 13, 2022}

\begin{abstract} 
The Zeno effect, in which repeated observation freezes the dynamics of a quantum system, stands as an iconic oddity of quantum mechanics. When a measurement is unable to distinguish between states in a subspace, the dynamics within that subspace can be profoundly altered, leading to non-trivial behavior. Here we show that such a measurement can turn a non-interacting system with only single-qubit control into a two-- or multi--qubit entangling gate, which we call a Zeno gate. The gate works by imparting a geometric phase on the system, conditioned on it lying within a particular nonlocal subspace. We derive simple closed-form expressions for the gate fidelity under a number of non-idealities and show that the gate is viable for implementation in circuit and cavity QED systems. More specifically, we illustrate the functioning of the gate via dispersive readout in both the Markovian and non-Markovian readout regimes, and derive conditions for longitudinal readout to ideally realize the gate.
\end{abstract}

\maketitle

\section{Introduction}
\label{sec:introduction}

The quantum Zeno effect refers to the ability of a strong repeated measurement to freeze the dynamics of a quantum system. Historically, this has been presented as an interesting and perplexing property of quantum measurements.
In recent years, however, quantum Zeno dynamics \cite{misra1977zeno,facchi2008quantum} are increasingly considered for quantum control and Hamiltonian engineering in quantum technology \cite{schafer2014experimental, raimond2012quantum, Mirrahimi_2014, Guillaud_2019}. 
In particular, previous work has shown that the Zeno effect can theoretically transform a trivial quantum system into a universal quantum computer \cite{burgarth2014exponential}.
This result, however, relies on arbitrary pulse-shaping, which makes it more similar to analog computation and less amenable to error correction and standard compilation methods. 

In this work, we instantiate such ideas in a more explicit model of gate-based quantum computation, in close connection with a recent experiment \cite{blumenthal2021}.
In particular, we illustrate how the Zeno effect can be used to implement two-- or multi--qubit \textsc{cphase} gates, which we call Zeno gates. 
The gate relies only on the ability to unitarily control a single qutrit in a cavity with other qubits, with no qubit--qubit coupling. 
In contrast with many existing schemes to create entanglement via measurement \cite{Williams_2008, martin2017optimal, Martin_2019_FlorEntangle,Lewalle_2021,Lewalle_2020_feed}, the present scheme relies on a measurement's ability to divide a large system into smaller subspaces, so as to non-locally inhibit a unitary operation.
Strong measurement of the correct subspace turns this trivial, non-interacting system into one with an effective entangling Hamiltonian. 
This kind of implementation of controlled phase gates has also been used in an existing proposal for a Zeno-based quantum operations based on interaction free measurements \cite{huang2008interaction} as well as for gates based on Coulomb blockades \cite{isenhower2010demonstration}. 
However, in contrast to those realizations, our gate does not require auxiliary qubits or qubit--qubit interactions.
We show in detail how the required Zeno measurement can be implemented in circuit or cavity QED systems.

The Zeno gate acting on $N$ qubits is locally equivalent to a $N$-qubit Toffoli gate, and hence it is universal for quantum computation when combined with single-qubit operations \cite{shi2002both}. 
The Zeno effect implies working in the limit of infinite measurement strength:~When the measurement strength is finite the Zeno effect can fail, which reduces the gate fidelity. We examine the behavior for the $N=2$ Zeno gate under finite measurement strength in detail in this work.
The Zeno gate can be implemented via an un-monitored dissipative channel (rather than a true measurement). 
We show that if the open channel implementing the Zeno effect is monitored, the gate can be heralded, which yields a probabilistic gate of higher fidelity. Despite its finite error rate, this heralded implementation of the gate can also be used for efficient universal computation, for instance by growing cluster states and then performing adaptive measurements \cite{barrett2005efficient, Santos_2012}.

The Zeno dynamical system we consider here can be solved numerically, or even analytically in some cases, although the resulting expressions are cumbersome. Therefore, we first focus on simple models that reproduce the important features of the exact solutions in simple closed-form expressions, before going on to account for increasingly realistic dynamical complications.
In Sec.~\ref{sec:ZenoGate}, we introduce the basic scheme for the Zeno gate and explain its operation in an ideal setting. This section yields a simple expression for the gate unitary without any need of approximations.
In Sec.~\ref{sec:FiniteGamma}, we analyze the effect of finite measurement rate, which is a source of infidelity that applies to all possible implementations of the gate. We show that the infidelity can be accurately estimated using an effective single qubit model and also derive an analytic expression for the fidelity of the heralded gate. 
In Sec.~\ref{sec:FiniteChi}, we outline how the Zeno gate could be implemented in circuit or cavity QED systems where the measurement is realized by a dispersive readout.
We identify regimes where the measurement is accurately modeled with a Markovian approach, discuss the limitations of these regimes for the Zeno gate, and identify sources of gate infidelity in the dispersive measurement dynamics. 
In Sec.~\ref{sec_Numerics} we then move beyond analytic models and perform numerical simulation of the Zeno gate in non--Markovian regimes, where the effective measurement strength depends on the history and interplay of the system and cavity used for readout. 
In Sec.~\ref{sec-longitude} we present an alternative implementation for circuit or cavity QED systems in which the measurement is realized by longitudinal readout, and show that this may offer a pathway to improved Zeno gate performance compared to that using dispersive readout.
Conclusions are summarized in Sec.~\ref{sec:Conclusion}, along with discussion about the outlook for further improving Zeno gate performance.
A number of technical and pedagogical details are presented in the appendices. 

Overall, the primary features of this work include the placement and study of the quantum Zeno effect in realistic scenarios for superconducting qubits, together with an emphasis on its use for gate--based quantum computation, in the context of a recent experiment \cite{blumenthal2021}. Our analysis characterizing the differences between heralded (conditional) Zeno dynamics and their unheralded (unconditional, or purely dissipative) counterpart also add substantively to the literature.

\section{Two- and N-qubit Zeno gates}
\label{sec:ZenoGate}

\begin{figure}
\centering
\includegraphics[width = 0.45\textwidth]{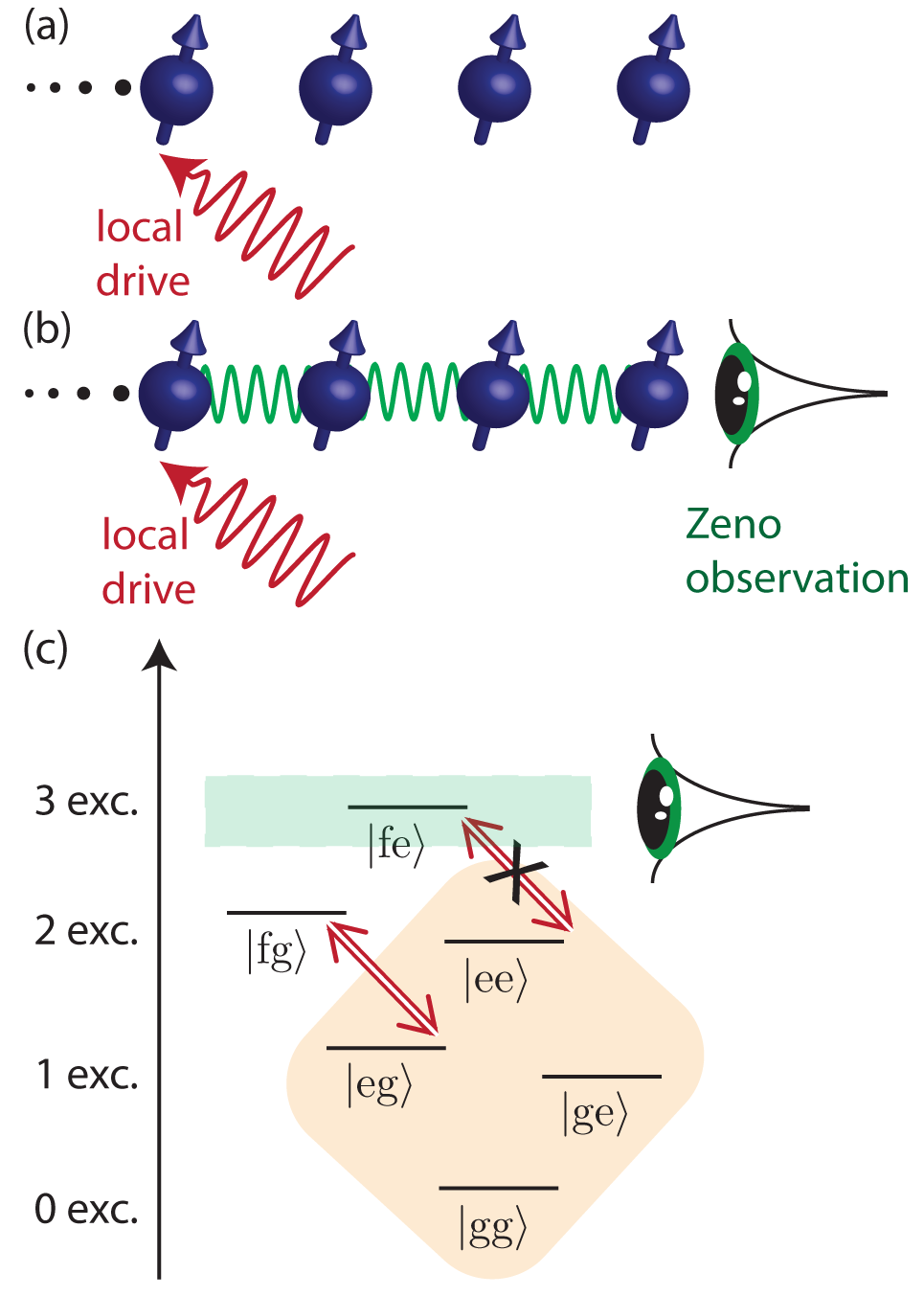}
\caption{(a) Isolated spins with local drives but no spin-spin interaction. (b) Measurement converts the application of local drives into implementation of an entangling gate. (c) Energy level diagram for a qutrit--qubit system, explaining the basic principle of the Zeno gate. 
The computational subspace ${\mathcal S}_Z$ is contained in the orange diamond. 
The green bar contains states in the three-excitation subspace, one of which is connected to the computational subspace by the Hamiltonian $\hat{H}$ (acting locally on the qutrit), which induces the transitions drawn in red. Frequent and repeated measurement of the projector $\hat{P}$ blocks population from the $|fe\rangle$ state, as indicated by the black cross. 
The combined action of $\hat{H}$ and $\hat{P}$ is a unitary but non-local operation on the qutrit--qubit system.
Under a full $2\pi$ rotation, a geometric phase is imparted on the $|eg\rangle$ level, which gives rise to an entangling gate within the computational subspace ${\mathcal S}_Z$.
}
\label{fig:LevelDiagram}
\end{figure}

We depict the system under consideration in Fig.~\ref{fig:LevelDiagram}(a). We consider one 3-level system (qutrit) and $N-1$ qubits, with no direct interactions. Our computational subspace is formed by the tensor product of the lowest two levels of the qutrit with the $N-1$ qubit space, which yields a $2^N$-dimensional computational space. 
We employ the naming convention used in superconducting circuits, in which the energy levels are labeled $|g\rangle$, $|e\rangle$ and $|f\rangle$ in order of increasing energy, and shall always place the qutrit first, i.e., on the left, in the tensor product state. 
For simplicity we start with a qutrit and a single qubit ($N=1$), as in the recent experiment of Blumenthal et al.~\cite{blumenthal2021}. 
Our goal will be to implement an entangling gate in the two-qubit computational subspace defined by the four states $\{|gg\rangle,|ge\rangle,|eg\rangle,|ee\rangle\}$, by driving the $|e\rangle\leftrightarrow |f\rangle$ transition of the qutrit. 
The system will always end in this computational subspace, either because we drive full $2\pi$ rotations or because the transition is blocked by the Zeno effect. 
We shall refer to this computational subspace as ${\mathcal S}_Z$.

\begin{figure}
\centering
\includegraphics[width = .6\columnwidth]{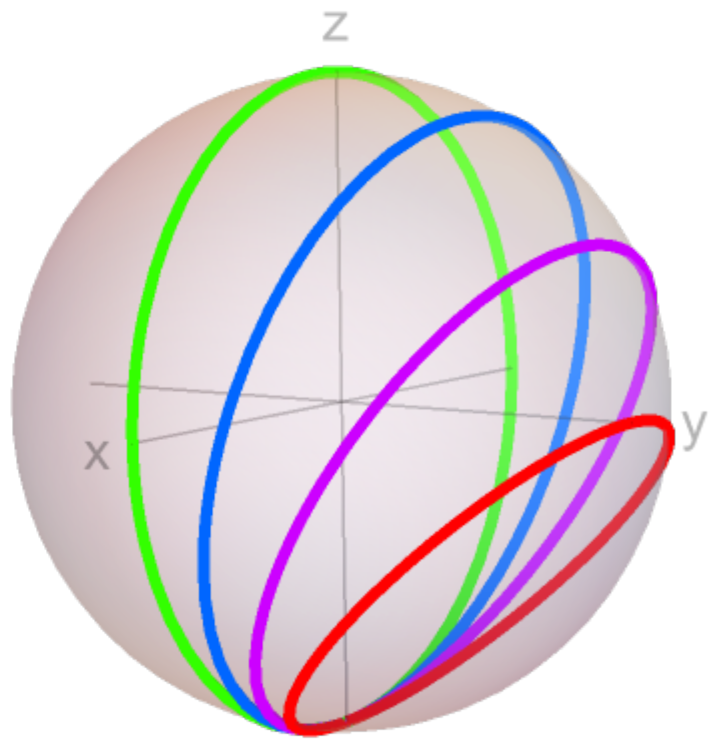}
\caption{
We illustrate different unitary operations which may be implemented on the $\ket{f} \leftrightarrow \ket{e}$ qutrit transition, in order to modify the geometric phase imparted by the Zeno gate. 
We show the operation \eqref{HRabi_3x2} in green, with other variants which subtend different solid angles on the $\ket{f} \leftrightarrow \ket{e}$ Bloch sphere in different colors. 
The Zeno gate is in general a \textsc{cphase} gate, which reduces to a \textsc{cz} gate in the special case \eqref{HRabi_3x2} that will be emphasized throughout the text for simplicity.
}
\label{fig:PhaseEnclose}
\end{figure}

We first consider the ideal scenario when utilizing the Zeno effect, in which one repeatedly applies an infinite strength projective measurement $\hat{P}$. Here that  measurement interrogates whether the qutrit--qubit system has exactly $3$ excitations, 
i.e.,
\begin{equation}
\hat{P} = 1-|fe\rangle\langle fe|.
\end{equation} 
This measurement does not distinguish between states in any other excitation subspace and in particular, it does not distinguish between states in the computational subspace $\mathcal{S}_Z$.
In the limit of frequent measurement of $\hat{P}$ considered here projection prevents the system from entering the three-excitation subspace
from the two or less excitation subspaces. 
We drive the $\ket{e}\,\leftrightarrow\,\ket{f}$ transition of the qutrit only, according to
\begin{align} \label{HRabi_3x2}
\hat{H} = \tfrac{i}{2}\,\Omega(|e\rangle\langle f|-|f\rangle\langle e|)\otimes \openone.
\end{align}
The above operations, when applied simultaneously, yield a unitary map that is locally equivalent to a \textsc{cz} gate when applied for a duration $T_G=2\pi/\Omega$. 
Fig.~\ref{fig:LevelDiagram}c depicts the basic concept. If the system starts in the computational subspace (orange diamond), then $\hat{H}$ alone drives the $|eg\rangle\leftrightarrow|fg\rangle$ and $|ee\rangle\leftrightarrow|fe\rangle$ transitions. 
However the latter transition is blocked by the Zeno effect under continuous measurement of $\hat{P}$, so $|ee\rangle$ is left untouched. 
As shown explicitly below, although  the $|eg\rangle$ component of the wave function is also mapped to itself under a full $2\pi$ rotation, it also picks up a geometric phase of $\pi$ in the process \cite{Burgarth2013non}.
This acquired phase is analogous and equivalent to the global phase acquired by a single qubit undergoing a $2\pi$ rotation. 
Thus the net operation applies a $\pi$ phase shift to the $|eg\rangle$ component of the state, while leaving all other components unchanged. 
This map is equivalent to a canonical \textsc{cz} gate when conjugated by a pair of $\pi$--pulses on the qubit before and after the gate.

The explicit form of the gate can be obtained by considering the action of $\hat{H}$ projected into the 2-excitation-or-less subspace 
via measurement of $\hat{P}$: \cite{facchi2008quantum} 
\begin{align}
\label{eq:DeriveHZeno}
\hat{H}_\text{Zeno} &= \hat{P} \hat{H} \hat{P}  \nonumber \\
&= \tfrac{i}{2} \Omega\, \hat{P} (|e\rangle\langle f|-|f\rangle\langle e|)\otimes(\underbrace{|g\rangle\langle g| + |e\rangle\langle e| 
}_{\openone}) \hat{P} \nonumber \\
&= \tfrac{i}{2}\Omega(|eg\rangle\langle fg|-|fg\rangle\langle eg|).
\end{align}
This Hamiltonian drives Rabi oscillations on the $|e\rangle\leftrightarrow|f\rangle$ levels of the qutrit conditioned on the qubit lying in the $|g\rangle$ state, which corresponds to the transition diagram of Fig.~\ref{fig:LevelDiagram}c. 
Since the Hamiltonian only acts non-trivially within a 2--qubit subspace, we can compute the corresponding unitary gate operation $\hat{U}_\text{Zeno} = \exp(-i\,t\, \hat{H}_\text{Zeno})$ on the 2--qubit subspace using Euler's formula 
\begin{align}
\label{eq:DeriveUZeno}
\hat{U}_\text{Zeno} &= (\openone-\hat{\Pi}_{eg,fg}) + \hat{\Pi}_{eg,fg}\cos(\Omega t/2) \\ \nonumber 
& \quad\quad + (|eg\rangle\langle fg| - |fg\rangle\langle eg|)\sin(\Omega t/2) \\ \nonumber
& = \openone-2\, \hat{\Pi}_{eg,fg} \quad\text{for}\quad t = 2\pi/\Omega,
\end{align}
where we have defined the operator $\hat{\Pi}_{eg,fg} \equiv |eg\rangle\langle eg| + |fg\rangle\langle fg|$ to project into the subspace affected by $\hat{H}_\text{Zeno}$ and we have chosen $t=T_G$ for the second equality. 

\erf{eq:DeriveUZeno} is the unitary operator for a controlled phase gate that applies a $\pi$ phase conditioned on occupying the $|eg\rangle$ level, as desired. 
Note that any other geometric phase may also be attained by subtending less than a hemisphere of the Bloch sphere when driving the $|e\rangle\leftrightarrow|f\rangle$ transition. 
We illustrate this in Fig.~\ref{fig:PhaseEnclose}. Specifically, the generalization of \eqref{HRabi_3x2} to
\be\begin{split} \label{HRabi_phase}
\hat{H} = \Omega\,\big\lbrace & \tfrac{i}{2}\,(|e\rangle\langle f|-|f\rangle\langle e|) \,\sin\phi \\ & \tfrac{1}{2}\,(\ket{e}\bra{e} - \ket{f}\bra{f})\,\cos\phi\big\rbrace\otimes \openone,
\end{split}\ee
allows for an arbitrary phase $\theta = \pi\,(\cos\phi + 1)$ to be imparted to the $\ket{eg}$ state, upon one period of the rotation generated by \eqref{HRabi_phase}. 
In other words, we may create evolution that is identical to a \textsc{cphase} gate, up to conjugation by local unitary operations \cite{Zhang_2012}.
The special case of \eqref{HRabi_3x2} generating a \textsc{cz} gate is recovered for $\phi = \pi/2$.
Note that the ability to impart an arbitrary phase allows one to generate infinitesimal gates, for example to generate Trotterized evolution of a desired Hamiltonian.

$\hat{U}_\text{Zeno}$ additionally generalizes straightforwardly to $N-1$ qubits interacting with a single qutrit. The result of this is an $N$--qubit unitary gate. 
This composite qutrit $\otimes$ ($N-1$) qubit system is our multi-level ``system'' and for convenience we shall also often refer to it as a qudit.
The relevant measurement operator and Hamiltonian are now
\begin{align}
\label{eq:ProjectorNQ}
&\hat{P}_N = \openone-|f e_2...e_N\rangle\langle f e_2...e_N| \\ \nonumber
&\hat{H}_N = \tfrac{i}{2}\Omega(|e\rangle\langle f|-|f\rangle\langle e|)\otimes \openone_2 \otimes ... \otimes \openone_N.
\end{align}
where we have only written the terms of $\hat{P}_N$ that are relevant to the dynamics. As before, we compute $\hat{H}_\text{Zeno}$ by conjugating $\hat{H}_N$ with $\hat{P}_N$. 
For notational compactness, we write $\hat{H}_\text{Zeno}$ by specifying its action only on the relevant subspace. For all $x_i \in \{g,e\}$ ($i = 2,...,N$) we have
\begin{align}
\hat{H}_{\text{Zeno},N}|g x_2...x_N\rangle &= 0 \\ \nonumber
\hat{H}_{\text{Zeno},N}|e x_2...x_N\rangle &= \left\{ \begin{array}{lr}
					0 	& : x_2...x_N=e...e\\
					-i|f x_2...x_N\rangle	&  \text{otherwise}
\end{array}
\right. \\ \nonumber
\hat{H}_{\text{Zeno},N}|f x_2...x_N\rangle &= \left\{ \begin{array}{lr}
					0 	& : x_2...x_N=e...e\\
					i|e x_2...x_N\rangle	&  \text{otherwise.}
\end{array}
\right. \\ \nonumber
\end{align}
Thus although $\hat{H}_N$ acts on a $3\times 2^{N-1}$ dimensional space, the projected Hamiltonian $\hat{H}_{\text{Zeno},N}$ acts non-trivially only on the $2^{N}$--dimensional subspace defined by $\ket{x_1\, x_2\,...\,x_N}$.
Although there is no explicit coupling within this qubit subspace, $\hat{H}_{\text{Zeno},N}$ can perform an entangling operation here, because of the non-local character of $\hat{P}_N$. By the same calculation as for \erf{eq:DeriveUZeno}, every component of the wave function of the form $|ex_1...x_N\rangle$ except $|ee...e\rangle$ picks up a $\pi$ phase, so that for $t=2\pi/\Omega$ we have 
\begin{equation}
\hat{U}_{\text{Zeno},N} = \exp(-i \,\hat{H}_{\text{Zeno},N}\, t) = \openone - 2\,\hat{\Pi}_{ex_1...x_N \neq ee...e}.
\end{equation}
Again, any phase between $0$ and $2\pi$ may be applied by subtending the corresponding solid angle on the Bloch sphere of each subspace.

$\hat{U}_{\text{Zeno},N}$ is locally equivalent to an $N$-body \textsc{cphase} gate, which in turn is locally equivalent to an $N$--qubit Toffoli gate \cite{shende2009cnot}, which together with local operations generates a universal gate set. To implement this $N$-body \textsc{cphase} gate, we simply apply another $2\pi$ rotation of $\hat{H}_N$, this time without measurement. Now all kets of the form $|e\,x_2...x_N\rangle$ acquire a $\pi$ phase, this time including $|ee...e\rangle$. All phases imprinted by this second operation cancel with those of $\hat{U}_{\text{Zeno},N}$ except that on $|ee...e\rangle$, resulting in an $N$--body \textsc{cphase} gate. Furthermore, one may imprint an arbitrary phase on $|ee...e\rangle$ by subtending different solid angles on the first and second application of $\hat{H}_N$. The full $N$--body $\pi$ \textsc{cphase} gate becomes a Toffoli gate when we conjugate any single qubit with a Hadamard gate before and after application of the \textsc{cphase} gate.

In the specific examples below, we will focus primarily on the two--qubit case for simplicity, with the understanding that the scheme can be scaled up straightforwardly. 
We will also often use the initial state $\lbrace \tfrac{1}{\sqrt{2}}\ket{e} + \tfrac{1}{\sqrt{2}}\ket{g} \rbrace ^{\otimes 2}$, because ideal implementation of the gate transforms this initially separable state into a maximally--entangled (Bell--like) state. 
Evolution from this initial state consequently offers a straightforward way to evaluate how much of the specifically \emph{non--local} work of the ideal gate is accomplished when using a realistically imperfect measurement channel to realize the projector $\hat{P}$.

This concludes our introduction of the ideal Zeno gate. 
We devote the remainder of the paper to characterizing the impacts of increasingly specific experimental imperfections on the Zeno gate.

\section{Implementation with Finite Measurement Strength}
\label{sec:FiniteGamma}

We now analyze the gate fidelity when the measurements are of finite strength, rather than the limit of infinitely rapid projective measurements. 
Measurements are always of finite strength in practice \cite{Guryanova2020}; our analysis in this section is thus more realistic, without yet focusing on the physics of a specific apparatus or implementation.
Finite measurement strength implies that our ability to ``Zeno block'' the unitary rotations at frequency $\Omega$ will be imperfect \cite{Snizhko2020Zeno,Kumar2020Zeno}.
We find that the first-order effect is leakage into the $|fe\rangle$ level blocked by measurement. 
A second-order effect appearing at weaker measurement strengths is a decrease in the amplitude of $|ee\rangle$ relative to the rest of the computational subspace. 
As explained below, this damping is caused by information acquired by a null measurement. The latter calculation also represents the gate fidelity when one uses the measurement signal to herald on successful implementation of the gate, which yields an improved fidelity and scaling of fidelity as a function of measurement rate.

The Zeno effect may be implemented either by sequential projective measurements or by a strong continuous measurement. 
Strictly speaking, our ability to Zeno block rests on the dissipation of information corresponding to a particular observable \cite{Harrington_2022}, a situation which is described by theories of open systems and continuous measurement \cite{Wiseman2009book, jacobs2014quantum}.
We now reconsider the Zeno gate from this viewpoint, which often constitutes a more realistic model compared to the projective measurements we appealed to above, while remaining tractable.
Schemes to implement non-trivial continuous measurements have been explored extensively in circuit QED systems \cite{blais2020circuit,Riste2013,Roch2014,roy2015remote,hacohen2016noncommuting,minev2018catch}. 
Continuous dissipation of information pertaining to an operator $\hat{P}$, combined with coherent evolution under Hamiltonian $\hat{H}$, is described by the master equation
\begin{align} \label{eq:ZenoMasterEquation}
\frac{d\hat{\rho}}{dt} &= -i[\hat{H},\hat{\rho}] + \Gamma\,\mathcal{D}[\hat{P}]\hat{\rho}
\end{align}
where $\Gamma$ is the measurement rate and $\mathcal{D}[\hat{P}] \hat{\rho} \equiv \hat{P}\,\hat{\rho}\, \hat{P}^\dag - (\hat{P}^\dag \hat{P}\, \hat{\rho} + \hat{\rho}\, \hat{P}^\dag \hat{P})/2$ is the standard Lindblad dissipator that models coupling to a Markovian bath \cite{gardiner2004noise}. 
Eq.~\eqref{eq:ZenoMasterEquation} leads to mixing of the state in measured subspaces due to averaging over all possible measurement outcomes weighted according to their probabilities. 
Further stochastic terms may be included to describe the evolution conditioned on a particular sequence of measurement outcomes \cite{Wiseman2009book, jacobs2014quantum}, however our ability to Zeno block a particular transition is well--described by the average evolution \eqref{eq:ZenoMasterEquation}. 
We will continue to refer to ``Zeno measurements'' below in a general sense, even when we use the average (i.e.,~Lindbladian) evolution for simplicity.\footnote{The effectiveness of \eqref{eq:ZenoMasterEquation} for describing the Zeno effect indicates that the Zeno effect may be understood as a feature arising from engineered dissipation \cite{Harrington_2022}, rather than measurement in the strictest sense. In the limit of large $\Gamma$, the dynamics initialized at an eigenstate (or eigenspace) of $\hat{P}$ are pinned there independent of whether or not we read the measurement outcome.}

\subsection{Unheralded Gate Fidelity}

Eq.~\eqref{eq:ZenoMasterEquation} can be solved exactly to extract a gate fidelity. 
The exact expressions are cumbersome, but we define $F_{\text{finite }\Gamma} = \langle \psi_0|\hat{U}_\text{Zeno}^\dag \,\hat{\rho}(t_\text{final})\, \hat{U}_\text{Zeno} |\psi_0\rangle$ with $\rho(0) \equiv |\psi_0\rangle \langle \psi_0|$, which is plotted for comparison with approximations derived below.

When the measurement rate $\Gamma$ is large but finite, the primary source of infidelity is population transfer into the $|fe\rangle$ state. 
Note that if the wave function collapses entirely to $|fe\rangle$ due to measurement, then gate fidelity is zero. 
We may consider dynamics only within the sub-space $\{|ee\rangle, |fe\rangle\}$ to estimate the probability of failed Zeno blocking.
Mapping this subspace to a single qubit $\tilde{\rho}$, with states $|\tilde{0}\rangle = |fe\rangle$ and $|\tilde{1} \rangle = |ee\rangle$, 
the master equation analogous to \erf{eq:ZenoMasterEquation} is
\begin{align} \label{eq:QubitZenoME}
\frac{d\tilde{\rho}}{dt} &= -i [\tilde{H}, \tilde{\rho}] + \Gamma\, \mathcal{D}[|\tilde{0}\rangle\langle \tilde{0}|]\tilde{\rho} \quad\text{with}\quad \tilde{H} = \Omega \frac{\tilde{\sigma}_y}{2}.
\end{align}
An exact solution is readily available (see Appendix \ref{sec:CalcDetails}). If the qubit is initialized in $|\tilde{1}\rangle$ and allowed to undergo a full $2\pi$ rotation, requiring $t=2\pi/\Omega$, then the probability to find the qubit in the leakage state $|\tilde{0}\rangle$ can be shown to be approximately
\begin{align} \label{eq:QubitZenoSol}
\tilde{P}_0 &\approx \frac{1-e^{-4\pi \Omega/\Gamma}}{2},
\end{align}
where we have dropped terms of order $\Omega^2$ and terms that are exponentially small in $\Gamma$ (see Appendix \ref{sec:CalcDetails} for details). This is consistent with the known result for blocking a superconducting qubit energy transition using the Zeno effect~\cite{gambetta2008trajectory}.

\erf{eq:QubitZenoSol} gives the contribution to the probability of failure of the Zeno gate due to the finite measurement strength allowing leakage from the $|ee\rangle$ state to a state outside the computational basis ($|fe\rangle$). To derive the corresponding probability of failure for the full Zeno gate, we simply multiply this by the probability to find the system in $|ee\rangle$, i.e.,~$\rho_{ee,ee}$. 
We define the first-order estimate of the gate fidelity $F_{\text{finite }\Gamma}^{(1)}$ by assuming that if the Zeno projection succeeds in preventing a transition to $|fe\rangle$, then the fidelity is $1$, while it is zero if the state is projected onto the state $|fe\rangle$, with intermediate values deriving from partial leakage. Using the probabilities derived above yields
\begin{align} \label{eq:FidFiniteGamma}
F_{\text{finite } \Gamma}^{(1)} &\equiv 1-\rho_{ee,ee}(0){\tilde P}_0
\end{align}
defined with resepct to a fixed initial state $\hat{\rho}(0)$.

\erf{eq:FidFiniteGamma} can be calculated using the exact, but rather lengthy, solution for ${\tilde P}_0 = 1 - \tilde{\rho}_{11}$ (again given in Appendix \ref{sec:CalcDetails}).
We can use the approximate expression for ${\tilde P}_0$ given in \erf{eq:QubitZenoSol} to make a simpler first-order estimate of $F_{\text{finite } \Gamma}$, namely
\begin{align} \label{eq:FidFiniteGamma_1}
F_{\text{finite } \Gamma}^{(1)} 
&\approx 1-\rho_{ee,ee}(0)\frac{1-e^{-4\pi \Omega/\Gamma}}{2}.
\end{align}
Both the exact form for $F_{\text{finite } \Gamma}$ and these first-order estimates yield unit fidelity in the ideal Zeno limit, i.e., when $\Gamma/\Omega \rightarrow \infty$, as required. 

\erfs{eq:FidFiniteGamma} and \erf{eq:FidFiniteGamma_1} represent the gate fidelity as a function of the initial state. An equally useful figure of merit is gate fidelity averaged over all pure input states, which we denote $\bar{F}$. In appendix \ref{sec:Haar}, 
we show that if the initial state satisfies $\rho_{ee,ee}(0)=1/4$, then the average fidelity $\bar{F}$ is identically equal to \erf{eq:FidFiniteGamma_1}.   Thus a useful test case for evaluation of the gate fidelity is the equal superposition state in the computational subspace, i.e., $(|gg\rangle+|ge\rangle+|eg\rangle+|ee\rangle)/2 = \lbrace \tfrac{1}{\sqrt{2}}\ket{e} + \tfrac{1}{\sqrt{2}}\ket{g}\rbrace^{\otimes 2}$. 
For this state, \erf{eq:FidFiniteGamma_1} yields 
\begin{align} \label{eq:FidBarFiniteGamma_equal}
    \bar{F}_{\text{finite }\Gamma}^{(1)} &\approx \frac{1}{8}\left(7+e^{-4\pi\Omega/\Gamma}\right).
\end{align}

\begin{figure}
\centering
\includegraphics[width = \columnwidth]{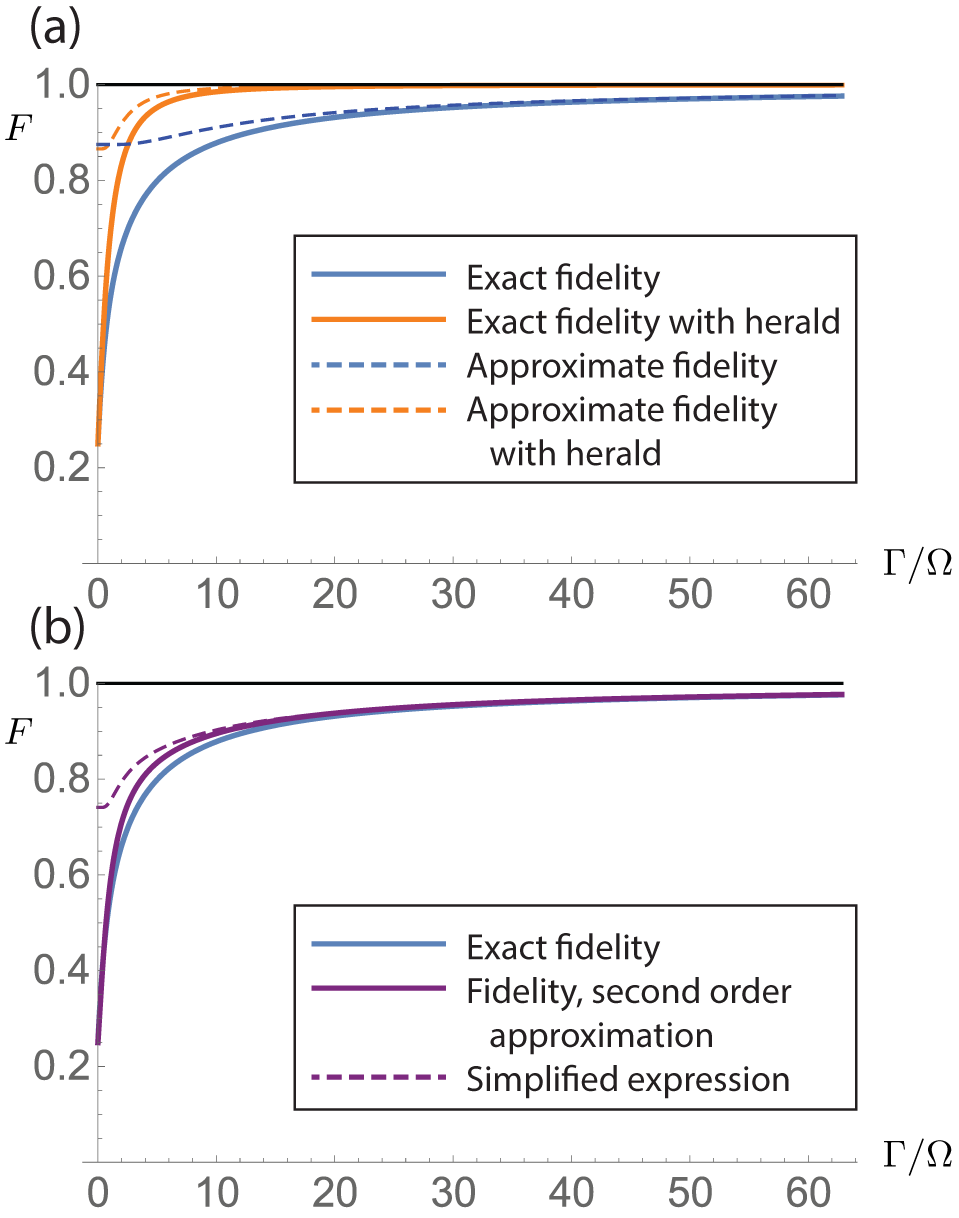}
\caption{
Zeno gate fidelity and approximations to this as a function of the unitless ratio $\Gamma/\Omega$.
The gate fidelity is calculated using the initial state $ (|gg\rangle+|ge\rangle+|eg\rangle+|ee\rangle)/2$ (see text for rationale for this choice). 
(a) Fidelities for the unheralded (blue) and heralded (orange) gate. Solid blue and orange lines indicate the exact results obtained by solutions of \erf{eq:ZenoMasterEquation} 
and \erf{eq:HNoJump} respectively, while the blue and orange dashed lines respectively plot the approximate expressions $F_{\text{finite }\Gamma}^{(1)}$ (\erf{eq:FidFiniteGamma_1}) and $F_\text{herald}$ (\erf{eq:FHerald}). (b) Comparison of the exact fidelity (blue line, same as in (a)) with second-order approximations to it. The solid purple line plots $F_{\text{finite }\Gamma}^{(2)}$ using the exact solutions in \erf{eq:FSecondOrder}, while the dashed purple line plots \erf{eq:FSecondOrderApprox}, where we have dropped some terms to yield a concise expression.}
\label{fig:FiniteGamma}
\end{figure}

Fig.~\ref{fig:FiniteGamma}a shows a comparison between the exact Zeno gate fidelity (blue solid line, calculated from \erf{eq:ZenoMasterEquation} using the exact form of ${\tilde P}_0$ given in Appendix A) and the average fidelity estimated using \erf{eq:FidFiniteGamma_1} (blue dashed line), as a function of the ratio $\Gamma/\Omega$. 
These calculations were performed with the initial equal superposition state for the reasons given above.
The agreement is good for large values of $\Gamma/\Omega$, i.e., high measurement rates. At smaller values of $\Gamma/\Omega$, our assumption in defining \erf{eq:FidFiniteGamma} that we have unit gate fidelity conditioned on success of the Zeno measurement, breaks down. Furthermore, even at large values of $\Gamma/\Omega$, convergence of this gate fidelity to unity is quite slow.

\subsection{Heralded Gate Fidelity} 

The results above are based on the average (or unconditional) dissipative evolution, which completely neglects knowledge of measurement outcomes specific to a given run.
We move beyond this by calculating the gate fidelity post--selected on realizations in which the measurement outcome corresponds precisely to this exclusion from $\ket{fe}$. 
We find that this post--selected fidelity converges to unity much faster than the unconditional fidelity calculated above. 
This intuitively indicates that the simpler unconditional evolution \eqref{eq:ZenoMasterEquation} may effectively be used to lower bound the Zeno gate's performance, but that detection and use of the conditional evolution or post--selection allow us to identify the best runs, in which the fidelity is much improved. 
In other words, our gate works much better if we are able to use our measurement outcomes to restrict our attention to runs of the experiment in which we do not detect escape to $\ket{fe}$.
While we do \emph{not} explicitly consider the additional possibility of feedback control for continuous error correction in the present work, it appears natural to suppose that the Zeno gate fidelity could be further improved via such methods.

To evaluate the fidelity with post--selection, we perform an unraveling of \erf{eq:ZenoMasterEquation} and compute the resulting quantum trajectories. There exist many equivalent ways to unravel \erf{eq:ZenoMasterEquation}, which represent different physical implementations of the measurement \cite{Wiseman2009book}. For calculational ease, we here assume a quantum jump model, in which the $|fe\rangle$ state undergoes decay to a continuum of states that we continuously monitor. If we never register population in these auxiliary states, then the system evolves under the following non-Hermitian Hamiltonian 
\begin{align} \label{eq:HNoJump}
    \hat{H}_\text{no jump} &= \tfrac{i}{2}\,\Omega(|e\rangle\langle f|-|f\rangle\langle e|)\otimes \openone + \tfrac{i}{2}\,\Gamma |fe\rangle \langle fe|
\end{align}
Note that one propagates the post selected state under $\hat{H}_\text{no jump}$ using only the Schr\"odinger equation, not the master equation \erf{eq:ZenoMasterEquation}. One can recover \erf{eq:ZenoMasterEquation} by evolving the state with \erf{eq:HNoJump} between quantum jumps to $|fe\rangle$ randomly applied according to the correct statistics, and then averaging over the resulting quantum jump trajectories.

The Schr\"odinger equation using \erf{eq:HNoJump} may be solved exactly. The solution is algebraically almost identical to that of \erf{eq:QubitZenoME} (see Appendix \ref{sec:CalcDetails}). Consequently, similar simplifications may be made to this solution by neglecting related terms. The main difference is that we solve now for a wave function instead of a density matrix. Setting $T_G=2\pi/\Omega$, we find
\begin{align} \label{eq:HNoJumpSol}
    \psi_{gg}(t) &= \psi_{gg}(0) \\ \nonumber
    \psi_{ge}(t) &= \psi_{ge}(0) \\ \nonumber
    \psi_{eg}(t) &= -\psi_{eg}(0) \\ \nonumber
    \psi_{ee}(t) &\approx e^{-\pi \Omega/\Gamma}\psi_{ee}(0) 
\end{align}
where we have also assumed that $\psi_{fe}(0)=0$, which yields $\psi_{fe}(t)=0$. One also finds a non-zero solution for $\psi_{fe}(t)$. However this undesired population can be eliminated by continuing measurement for a time that is large compared to $1/\Gamma$, which exponentially damps $\psi_{fe}$. Thus we take $\psi_{fe}$ to be zero.

The heralded gate fidelity is simply the overlap squared of the above state with the initial state under application of an ideal gate as described in Sec. \ref{sec:ZenoGate}. The only subtlety is that the post-selected wave function above must first be normalized, since evolution under a non-Hermitian Hamiltonian does not preserve the norm. The resulting heralded gate fidelity is
\begin{align} \label{eq:FHerald}
    F_\text{herald} &\approx \frac{1-|\psi_{ee}(0)|^2(1-e^{-\pi \Omega/\Gamma})}{\sqrt{1-|\psi_{ee}(0)|^2(1-e^{-2\pi \Omega/\Gamma})}},
\end{align}
where we write `$\approx$' because we have used the approximate solutions of Eqs.~\eqref{eq:HNoJump} and \eqref{eq:HNoJumpSol}.

We can use \erf {eq:FHerald} to compute a state-averaged fidelity $\bar{F}_\text{herald}$. However, due to the post-selection, $F_\text{herald}$ is non-linear in $|\psi\rangle$, and thus the usual methods of calculating the average fidelity analytically fail. 
In Appendix \ref{sec:Haar}, we compute the average fidelity by numerically integrating over the Haar measure and thereby show that $\bar{F}_\text{herald}$ is well approximated by the value of \erf{eq:FHerald} for the initial equal superposition state $(|gg\rangle+|ge\rangle+|eg\rangle+|ee\rangle)/2$, just as was the case for the unheralded gate in \eqref{eq:FidFiniteGamma_1}--\eqref{eq:FidBarFiniteGamma_equal}.

$F_\text{herald}$ plays a dual role. First, is its interpretation in the context of a heralded implementation of the gate. In Fig.~\ref{fig:FiniteGamma}a) we plot the heralded fidelity in orange, together with the unheralded fidelity $F_{\text{finite } \Gamma}$ (blue solid line) and its first order approximation $F_{\text{finite } \Gamma}^{(1)}$ (blue dashed line). 
We see that both the exact and first order fidelities converge to unity orders of magnitude more quickly with heralding. In this setting
$\bar{F}_{\text{finite }\Gamma}^{(1)}$ may be interpreted as the success probability, or the fraction of the time in which measurement indicates that no jump to $|fe\rangle$ occurred.

Secondly, \erf{eq:FHerald} provides a second-order correction to the unheralded fidelity. Since $F_\text{herald}$ explicitly removes the infidelity that remains in
$F_{\text{finite }\Gamma}^{(1)}$, we can compute the total fidelity by subtracting both of the corresponding infidelities, i.e., for finite measurement strength and for heralding. 
The result is
\begin{align} \label{eq:FSecondOrder}
    F_{\text{finite }\Gamma}^{(2)} &\equiv 1-(1-F_{\text{finite }\Gamma}^{(1)}) - (1-F_\text{herald}) \\ \label{eq:FSecondOrderApprox}
    &\approx \frac{1-|\psi_{ee}(0)|^2(1-e^{-\pi\Omega/\Gamma})}{\sqrt{1-|\psi_{ee}(0)|^2(1-e^{2\pi\Omega/\Gamma})}} \nonumber \\
    &\quad + \frac{|\psi_{ee}(0)|^2}{2}\left(1-e^{-4\pi\Omega/\Gamma} \right)
\end{align}
The infidelity estimated in $F_{\text{finite }\Gamma}^{(1)}$ has a clear physical interpretation of leakage into the $|fe\rangle$ state. Similarly, the infidelity calculated in $F_\text{herald}$ has a simple physical explanation, namely the indirect acquisition of information. Returning to \erf{eq:HNoJumpSol}, observe that the only deviation from an ideal gate is damping of the $\psi_{ee}$ component of the wave function. 
Only $\psi_{ee}$ population can lead to population in $|fe\rangle$, so if we do not measure $|fe\rangle$ population, then we can infer a lower likelihood to find the system in $|ee\rangle$. 
This damping is analogous to the case of an atom prepared in a superposition of its excited and ground states, which decays to its ground state even if it does not emit a photon \cite{FlorTeach}. 

We plot this second order corrected fidelity against the exact fidelity in Fig.~\ref{fig:FiniteGamma}b). We also include the fidelity calculated using the exact solutions to \erf{eq:ZenoMasterEquation} and \erf{eq:HNoJump}, which are omitted in the main text for brevity but are given in Appendix \ref{sec:CalcDetails}. 
The three fidelity curves agree well, indicating that by accounting for both the infidelity due to finite measurement strength and that due to the information gain (heralding) we have quantified the main sources of infidelity in an intuitive, closed-form expression.

As a final note, we remark that the full gate error in diamond norm can be bounded as $\mathcal{E}_\diamond < 38\;\Omega/\Gamma$~\cite{blumenthal2021,Burgarth2019generalized}.

\section{Implementation in Cavity QED and finite coupling effects}
\label{sec:FiniteChi}

We continue towards more realistic measurement dynamics:
We will presently describe how the Zeno gate can be implemented using dispersive measurements, which are common in superconducting qubit systems \cite{Blais2004, gambetta2008trajectory, Korotkov2016, blais2020circuit}.
This measurement technique leads to finite measurement strengths as just discussed, but also to further dynamics which are not strictly necessary for the Zeno gate, and are not captured by \erf{eq:ZenoMasterEquation}.
In the cavity QED implementation discussed below, we must not only consider infidelity due to leakage through our Zeno block, but also due to spurious damping (dephasing within the computational subspace $\mathcal{S}_Z$) and Stark shifts induced by measurement.
In analyzing these effects, we essentially consider one way that dynamics like \erf{eq:ZenoMasterEquation} are realized in practice.

\begin{figure}
\centering
\includegraphics[width = \columnwidth]{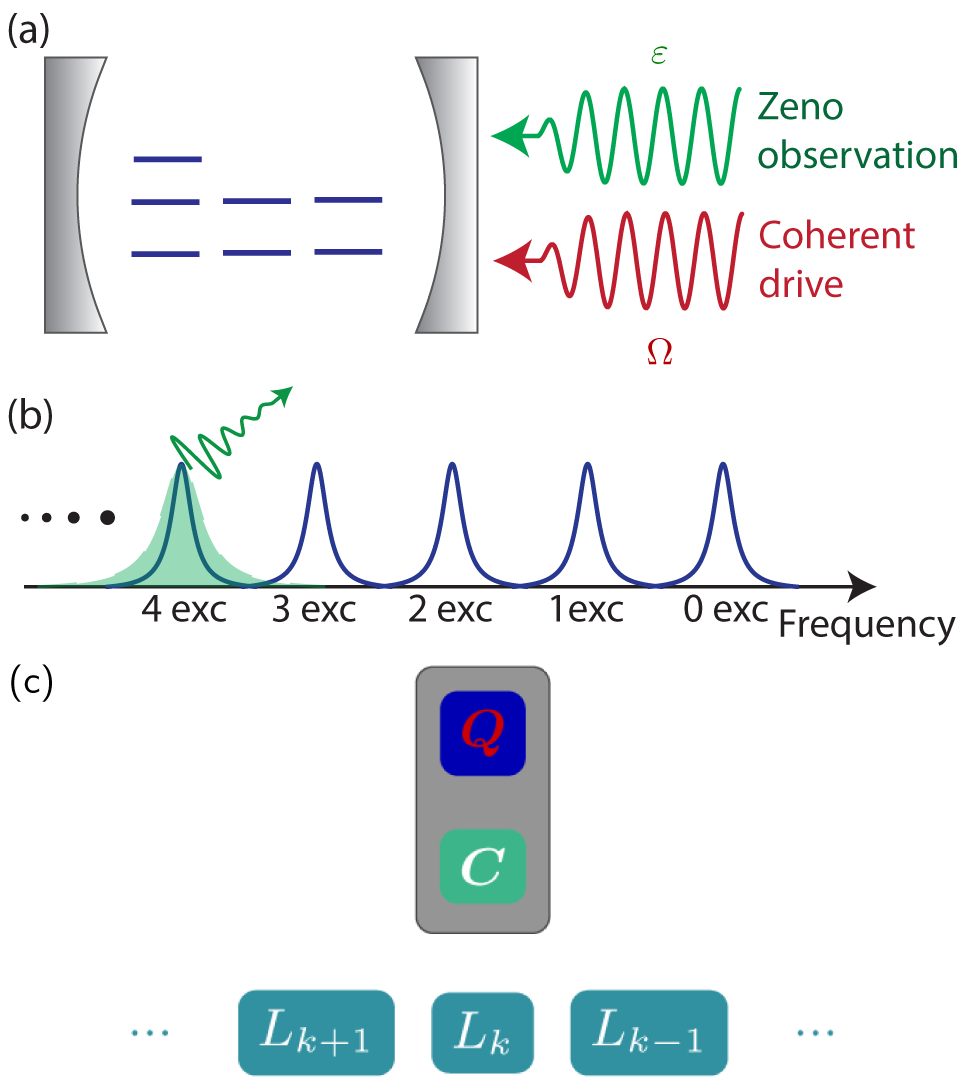}
 \caption{
 Cavity QED implementation of the $N$-qubit Zeno gate. a) Several transmons are coupled dispersively to a high-finesse cavity with leakage rate $\kappa$. For the $N-1$ transmons acting as qubits, we only show the lowest two levels. 
 The Zeno measurement $\hat{P}$ is implemented by the action of \eqref{eq_HDisp} on the qudit--cavity system, which includes unitary rotations $\boldsymbol{\Omega}$ on the qudit, dispersive coupling between the qudit and cavity, and a readout tone driven by a classical field of frequency $\omega_\varepsilon$ close to the bare cavity resonance $\omega_c$. 
 b) Cavity density of states as a function of frequency, plotted for all qubit states in the computational basis. This plot can also be interpreted as the intracavity field as a function of drive frequency. Due to the nature of the dispersive Hamiltonian, the effective cavity resonances cluster in groups of equal qubit excitation number, facilitating measurement of $\hat{P}$. 
 c) Schematic illustrating the leakage of the measurement tone into the external line, which may be directly monitored, with a structure reminiscent of that in composite quantum collision models \cite{ciccarello2021, Lorenzo2017}. In the present case of interest, the qudit (Q) and cavity mode (C) are isolated together, and interact via dispersive coupling according to \eqref{eq_HDisp}. Their only contact with the outside world is mediated through cavity decay into the external line (L). One can then describe the measurement process and/or decoherence occurring via this channel by envisaging a sequence of empty lines (i.e.,~temporal line modes $L_k$, initialized in vacuum) that arrive and interact with the qubit--cavity system at time $k$, each for a duration $\Delta t$. 
 }
\label{fig:cQEDImplementation}
\end{figure}

Dispersive readout provides a well--known way to implement a measurement of excitation number; it is realized by placing a qudit in a leaky cavity, as illustrated in Fig.~\ref{fig:cQEDImplementation}.  
Dispersive readout is based on a multi-level generalization of the Rabi Hamiltonian (i.e.,~a transverse coupling between the qudit and cavity mode)
\be \label{eq_Htransverse}
\hat{H}_\mathrm{tv} = \omega_c\,\hat{a}^\dag\hat{a}+\sum_j \omega_j \ket{j}\bra{j} +\sum_{j\neq\ell} g_{j\ell}(\hat{a}+\hat{a}^\dag)\ket{j}\bra{\ell}.  
\ee
States $\ket{j}$ are for the qudit, while $\hat{a}^\dag$ and $\hat{a}$ are cavity photon creation and annihilation operators, respectively. 

By first making the rotating wave approximation, and then the dispersive approximation (the latter assumes the cavity is far-detuned from the natural transition frequencies of the system, or more specifically that $|g_{\ell j}| \sqrt{\langle \hat{a}^\dag\hat{a}\rangle+1} \ll |\omega_\ell - \omega_j - \omega_c|$), one finds that (see \cite{blais2020circuit} and references therein)
\begin{subequations} \be 
\hat{H}_\mathrm{disp} = \sum_j\left(\omega_j + \lambda_j + \chi_j\,\hat{a}^\dag\hat{a} \right)\ket{j}\bra{j}, \quad\text{with}
\ee \be
\lambda_j \equiv \sum_\ell \frac{|g_{\ell j}|^2}{\omega_j-\omega_\ell -\omega_c},\quad\text{and} 
\ee \be 
\chi_j \equiv \sum_\ell\left( \frac{|g_{\ell j}|^2}{\omega_j - \omega_\ell - \omega_c} - \frac{|g_{\ell j}|^2}{\omega_\ell-\omega_j - \omega_c}\right). 
\ee \end{subequations} 
The state-dependent dispersive shifts $\chi_j$ are key to readout going forward. 
Note that because the photon number $\hat{a}^\dag\hat{a}$ is directly tied to the action of these dispersive frequency shifts, high photon numbers can increase measurement strength, but eventually break the validity of the dispersive approximation.
We include a coherent drive $\varepsilon(\hat{a}\,e^{i\,\omega_\varepsilon\,t+i\,\phi} + \hat{a}^\dag\,e^{-i\,\omega_\varepsilon\,t-i\,\phi})$ detuned from the bare cavity frequency by $\Delta^c_\varepsilon = \omega_c - \omega_\varepsilon$ that supplies amplitude to the readout mode at rate $\varepsilon$, as well as Rabi rotations $\boldsymbol{\Omega}$ which can be implemented by tones on--resonance with the qudit transitions. The full Hamiltonian within the dispersive approximation can then be re-written
\be
\label{eq_HDisp}
\hat{H}_\text{q--c} = \tfrac{1}{2}\,\boldsymbol{\Omega}\cdot\hat{\boldsymbol{\sigma}}+ \left( \Delta^c_\varepsilon + \boldsymbol{\chi}\cdot\hat{\boldsymbol{\Pi}} \right) \hat{a}^\dag\hat{a} + \varepsilon(\hat{a}^\dag\,e^{i\phi}+\hat{a}\,e^{-i\phi}),
\ee
where we have gone into the interaction picture with the cavity degrees of freedom rotating at $\omega_\varepsilon$, and eliminated the bare qudit terms.
We have shorthanded the qudit operators $\hat{\boldsymbol{\sigma}}$ (e.g.~such that $\tfrac{1}{2}\,\boldsymbol{\Omega}\cdot\hat{\boldsymbol{\sigma}} = \tfrac{1}{2}\Omega_{j\ell}\ket{j}\bra{\ell}$, or a set of Gell--Mann matrices) and the qudit energy--eigenstate projectors $\hat{\boldsymbol{\Pi}}$  (such that $\boldsymbol{\chi}\cdot\hat{\boldsymbol{\Pi}} = \sum_j \chi_j\,\hat{\Pi}_j$) for concision. 
The choice of driving quadrature, determined by $\phi$, has no impact on the dynamics of interest, and will be chosen as is convenient at points below.

In order to use the above for measurement, it is necessary that the cavity be driven, but also damped, such that light exits the cavity into a readout line where it is either dissipated or detected. 
We assume Lindbladian dissipation of the cavity 
as per
\begin{align} \label{eq:OpenCavME}
    \frac{d\hat{\rho}}{dt} = -i[\hat{H}_\text{q--c},\hat{\rho}] + \kappa \, \mathcal{D}[\hat{a}]\hat{\rho}, 
\end{align}
where $\kappa$ is the cavity linewidth. See Fig.~\ref{fig:cQEDImplementation}.
Below we shall discuss the non-Markovian (NM) dynamics resulting from unitary qudit rotations and comparatively slow cavity decay, which cause the cavity and qudit to remain entangled and ``remember'' each other's dynamics over time. 
However, we do \emph{not} at any point consider the further possible effects of NM dynamics due to extended correlations between the cavity and the external line/environment.

In our upcoming discussions, we shall weigh the impact of two different timescales that affect the Zeno gate, namely the gate time (which is set by $\Omega= |{\bf \Omega}|$) and the cavity decay rate $\kappa$. 
We begin by considering measurement \emph{in the absence} of coherent qudit rotations (i.e.,~with $\boldsymbol{\Omega} = 0$), drawing on existing exact solutions for this case for arbitrary $\kappa$ \cite{Korotkov2016}.

\subsection{Dispersive measurement alone \label{sec-CA-model}}

Within the context of the dispersive approximation, and with $\boldsymbol{\Omega} = 0$, coherent cavity states will remain coherent \cite{Korotkov2016, blais2020circuit, steinmetz2021}. 
This allows us to write a joint qudit--cavity state of the form 
\begin{align} \label{eq:rhoPolaron}
    \rho(t) &= \sum_{j,\ell} \rho_{j \ell} |j\rangle\langle \ell|\otimes |\alpha_j(t)\rangle\langle\alpha_\ell(t)|,
\end{align}
where the sum is taken over all relevant qudit states ($j = gg,ge,eg...$). 
The evolution of the coherent state amplitudes $\alpha_j(t)$ may be solved for using the Heisenberg equations of motion and input-output theory to handle the dissipator \cite{clerk2010noise, Korotkov2016}. This results in the equation of motion for the field operator evolution
\begin{subequations}
\be 
\dot{a} = i[\hat{H}_\text{q--c},\hat{a}] - \tfrac{1}{2}\kappa\,\hat{a} = -i\,\varepsilon\,e^{i\phi} - \left(i\Delta_j + \tfrac{1}{2}\kappa\right) \hat{a},
\ee
where $\varepsilon$ is again the rate at which amplitude is supplied to the measurement tone, $\kappa$ is the cavity decay rate, and $\Delta_j = \Delta^c_\varepsilon + \chi_j$ is a detuning including the frequency difference between the bare cavity resonance and the drive $\varepsilon$ as well as the level--dependent dispersive shift $\chi_j$. 
As with \eqref{eq_HDisp}, we work in the frame rotating at the frequency $\omega_\varepsilon$ of the drive tone.
The associated coherent state amplitudes then vary as (with $\phi = \pi/2$)
\be 
\bra{\alpha_j}\dot{a}\ket{\alpha_j} = \dot{\alpha}_j = \varepsilon - \left(i\Delta_j + \tfrac{1}{2}\kappa\right) \alpha_j. 
\ee
\end{subequations}
This equation admits the analytic solution
\begin{subequations}
\be \label{alpha-solution}
\alpha_j(t) = \left(\alpha_j(0)-\frac{2\,\varepsilon}{\kappa+2i\,\Delta_j}\right)e^{-\left(i\,\Delta_j+\tfrac{1}{2}\kappa\right)t}+\frac{2\,\varepsilon}{\kappa+2i\,\Delta_j}
,\ee
which decays towards the steady state 
\be \label{cavss} 
\bar{\alpha}_j = \frac{2\varepsilon}{\kappa + 2i\,\Delta_j} = \frac{2\varepsilon}{\sqrt{\kappa^2 + 4\Delta_j^2}} \exp\left[ i\,\arctan\left( \frac{2\Delta_j}{\kappa} \right) \right]
\ee \end{subequations}
on a timescale $\kappa^{-1}$. 
It will additionally be important to consider phase evolution $\dot{\varphi}$ associated with each coherent state $e^{i\varphi_j}\ket{\alpha_j}$ \cite{Korotkov2016, Cross_2015, paik2016experimental}. This phase evolves according to
\be \label{outerphase}
\dot{\varphi}_{j} =  -\varepsilon\,\Re\left[e^{-i\phi}\,\alpha_j \right], 
\ee 
such that its accumulation may be straightforwardly computed from \eqref{alpha-solution}. 
Note that the driving quadrature $\phi$ and resonator--induced phases $\varphi$ are distinct physical quantities.
Taken together, these expressions lead to qudit--cavity dynamics expressed as linear first order differential equations.
Those equations are uncoupled for $\boldsymbol{\Omega} = 0$, but may become coupled once the Rabi drive is turned on, which directly threatens the legitimacy of the ansatz \eqref{eq:rhoPolaron}.

The resulting Lindbladian qudit--cavity dynamics are characterized by the effective dephasing rate between any two states
\be \label{meas-strength}
\Gamma_{j \ell} \equiv \kappa\left(\tfrac{1}{2}|\alpha_j|^2 + \tfrac{1}{2}|\alpha_\ell|^2 - \Re\left[\alpha_j\,\alpha_\ell^\ast \right] \right), 
\ee 
and relative phase rotations at a rate
\be \label{ac-stark} 
\Upsilon_{j\ell} = \kappa\, \Im[\alpha_j\,\alpha_\ell^\ast] + \varepsilon\,\Im\left[\alpha_\ell - \alpha_j \right], 
\ee
(summing contributions due to dissipation and \eqref{outerphase}).
See Appendix \ref{app-qubitdisp} for details. 
We take \eqref{meas-strength} to be an expression for the effective strength of the measurement (i.e.,~this is the rate at which information enabling the qudit states $\ket{j}$ and $\ket{\ell}$ to be distinguished dissipates out of the cavity, thereby becoming available to an external observer), while \eqref{ac-stark} contains both the the AC Stark shift and resonator--induced phase (RIP) \cite{Korotkov2016, Cross_2015, paik2016experimental}. The second term in $\Upsilon$ is itself the basis of an entangling gate (RIP gate).

In the text below, we will often refer to a ``coherent--state analytical'' (CA) model; this refers specifically to the model just introduced, using coherent cavity states (i.e.,~based on the ansatz \eqref{eq:rhoPolaron}).

\subsection{Measurement of a Specific Excitation Number}

We can study the unconditional dynamics given by the CA model (still with $\boldsymbol{\Omega} = 0$) to understand how to tune a dispersive measurement for use in the Zeno gate, keeping in mind that any finite efficiency detection will \emph{improve} performance compared to the average (unconditional or unheralded) evolution. %
We will emphasize the situation of a qutrit $\otimes$ qubit $(N=2)$ for the purposes of analysis and numerical examples in the remainder of the manuscript. 

Several parameters need to be chosen or engineered to implement a good measurement operator $\hat{P}$ for the Zeno gate, including $\omega_\varepsilon$, $\kappa$, and the dispersive shifts $\chi_j$. 
We wish to choose these parameters in a way that leads to a measurement with two key properties:
\begin{enumerate} 
\item We wish to measure the transition to $\ket{fe}$ very strongly. This requires that $\Gamma_{ij}$ be as large as possible for any transition with $i = fe$ or $j = fe$. 
\item We wish to maintain coherent evolution in the subspaces with lower excitation number, which means that we want $\Gamma_{ij}$ \emph{as small as possible} for all transitions $i,j \neq fe$ (so that our Zeno measurement does not accidentally decohere the computational subspace $\mathcal{S}_Z$). 
\end{enumerate}
In combination, these properties guarantee a dissipation or measurement that divides our space by excitation number between $\lbrace \ket{fe} \rbrace$ and $\lbrace \ket{fg}, \ket{ee}, \ket{eg}, \ket{ge}, \ket{gg} \rbrace$. In other words, these requirements ensure behavior mimicking the key properties of the projector $\hat{P}$, thereby creating the Zeno effect we want for our gate.
We may understand how to satisfy both properties simultaneously by evaluating the steady state measurement strength (i.e.,~by evaluating \eqref{meas-strength} for cavity states \eqref{cavss}). 
The steady--state measurement strength $\bar{\Gamma}_{ij}$ for any given transition has two peaks when plotted as a function of drive tone frequency $\omega_\varepsilon$, characterized by $\Delta_i = 0 = \Delta_j$, i.e.,~for $\omega_\varepsilon = \omega_c + \chi_i$ and $\omega_\varepsilon = \omega_c + \chi_j$. 
These peaks are Lorentzian, and the decay rate $\kappa$ sets the (line)width. 
It follows that in order to maximize our measurement strength for transitions to $\ket{fe}$, we wish drive our measurement tone at the frequency $\omega_\varepsilon = \omega_c + \chi_{fe}$ (i.e.,~we choose $\Delta_{fe} = 0$). 
In order to also satisfy our second condition, we require that this peak corresponding to $\ket{fe}$ that we select be well separated from all the others: This can be accomplished by choosing $\kappa$ \emph{smaller} (preferably by a factor of \emph{at least} two or three) than the difference $\Delta \chi = |\chi_{fe} -\chi_j|$ where $\chi_j$ corresponds to the nearest neighboring peak (ostensibly $\chi_{ee}$ or $\chi_{fg}$). 
An example is shown in panels (a) and (b) of Fig.~\ref{fig-linestrength}.

\subsection{Timescales for the Zeno Gate}
\label{sec:constraints}

\begin{figure}
\includegraphics[width = \columnwidth]{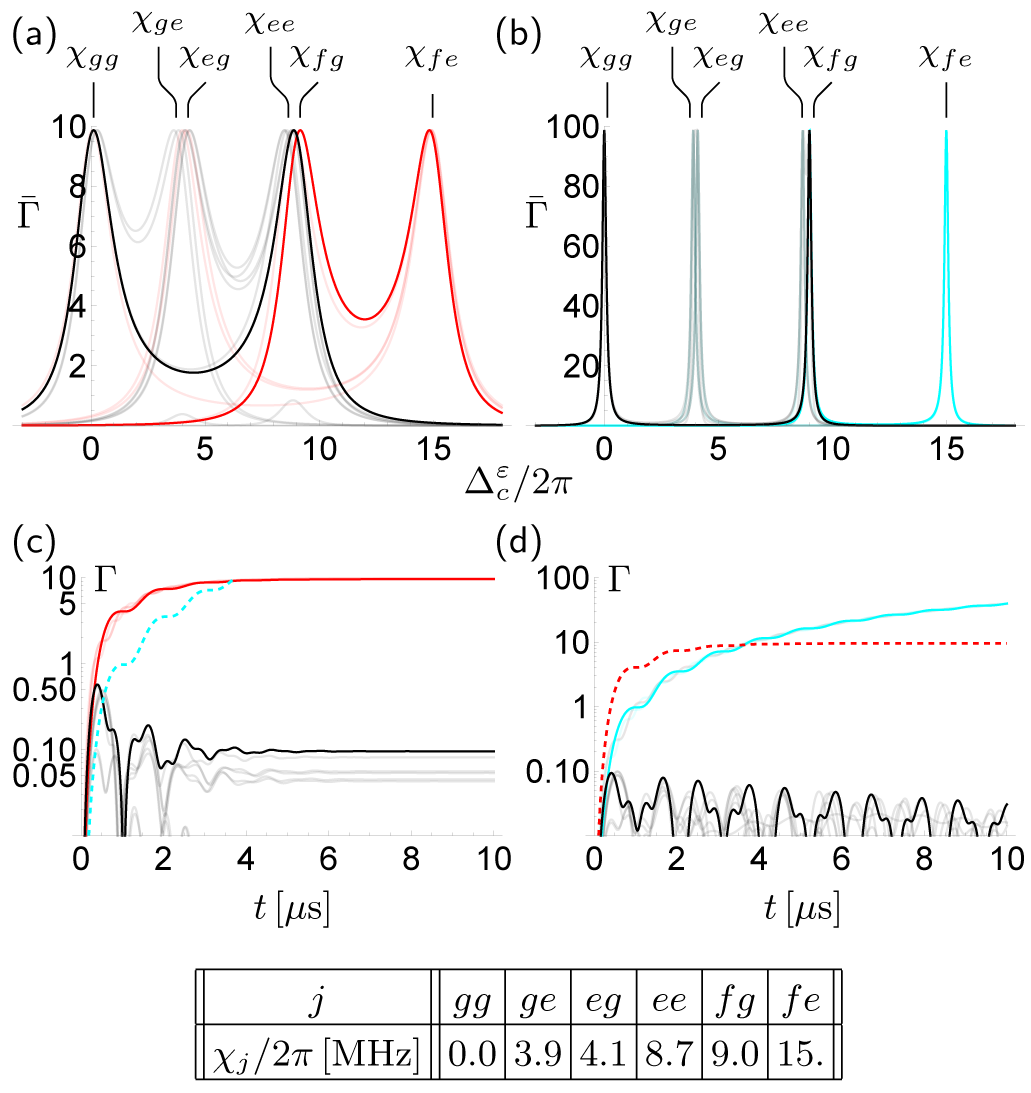}
    \caption{
   Analysis of the basic dispersive measurement requirements to implement an $N=2$ Zeno gate, based on the CA model.
   Panels (a) and (b) plot the steady--state measurement strength as a function of the readout measurement drive frequency $\omega_\varepsilon$ (shown in terms of the magnitude of the relative frequency $\Delta^\varepsilon_c =-\Delta^c_\varepsilon = \omega_\varepsilon -\omega_c$), for two different values of cavity linewidth, $\kappa = 2\,\mathrm{MHz}$ in (a) and $\kappa = 0.2\,\mathrm{MHz}$ in (b). 
   The measurement rate for transitions to/from $j=fe$ are plotted in red (a) or cyan (b), while transitions within $\mathcal{S}_Z$ are shown in black. 
   Panels (c) and (d), plot the ring-up of the measurement strength as a function of time, given the cavity initially in vacuum, with $\kappa = 2\,\mathrm{MHz}$ in (c) and $\kappa = 0.2\,\mathrm{MHz}$ in (d), evaluated for an initially empty cavity, i.e., with $\alpha_0 = 0$. 
   In other words, (c) uses the same $\kappa$ as (a), while (d) uses the same $\kappa$ as (b); the ring-up using the ``other'' value is plotted in each of panels (c) and (d) as a dashed line in its contrasting color for easy comparison.
   The measurement tone is driven at amplitude $\varepsilon/2\pi = 1\,\mathrm{MHz}$ in all panels and the values of $\chi$ employed are shown in the table below panels (c) and (d). The specific values of $\chi$ are arbitrarily chosen, but are all of a scale that is realistic on existing transmon systems. 
    }
    \label{fig-linestrength}
\end{figure}

Panels (c) and (d) of Fig.~\ref{fig-linestrength} illustrate a consequence imposed by the measurement regime we have just described.  
Clearly large $\chi$ values are desirable to maximize $\Delta \chi$, but these dispersive shifts cannot be made arbitrarily large in practice. 
This means that once we have maximized $\Delta \chi$ to within attainable constraints, the linewidth $\kappa$ is constrained in turn.  
Notice that the solutions \eqref{alpha-solution} decay towards their steady state value (and steady state average photon number $|\bar{\alpha}|^2$) at rate $\kappa$. 
The CA model thus tells us that while a smaller $\kappa$ value allows us to resolve a particular transition with a strong measurement rate, it also slows the speed at which the cavity can respond to other dynamics. 
In other words, $\kappa^{-1}$ sets the timescale on which an initially separable qudit--cavity state $(\sum_j c_j \ket{j})\otimes \ket{\alpha_0}$ evolves towards an entangled steady state of the form $\sum_j c_j' \ket{j}\ket{\bar{\alpha}_j}$ offering a stable measurement strength of the qudit via the cavity pointer states.
Thus, panels (c) and (d) of Fig.~\ref{fig-linestrength} illustrate the trade-off between these two features, with the  cavity response clearly slower for the smaller $\kappa$ value (d) even though the eventual steady--state measurement is stronger.\footnote{We suggest that this slowing of the cavity dynamics moving from the initial state towards a steady state may be at least partially mitigated by shaping $\varepsilon(t)$ instead of using a constant value; in particular, one could drive the cavity harder at the start of the measurement interval to overcome the slow ring--up time, and then lower $\varepsilon$ to the level necessary to maintain the desired steady state cavity population.}

We now begin considering $\boldsymbol{\Omega} \neq 0$, which is the last ingredient required for the Zeno gate.
Rabi rotations on the qudit introduce a significant element with regards to the measurement itself (assuming they do not commute with the monitored observable), in that the qudit dynamics occurring independent of the measurement tone will now generically ``drag'' the cavity pointer states behind \cite{gambetta2008trajectory}. 
This can distort the cavity states so as to render the ansatz \eqref{eq:rhoPolaron} that is the basis of the CA model approximate at best, and wholly invalid at worst, depending on one's choice of parameters.
It is helpful at this point to consider two opposing regimes: 
\begin{enumerate}
\item In the limit $\kappa \gg \Omega$ effects such as distortion of the cavity pointer states are negligible (one can imagine that the dynamics of the cavity pointer states are rapid enough to adiabatically follow the comparatively slow Rabi rotations); this is the usual Markovian regime suitable for continuous measurement \cite{Murch:2013ur}. 
In this case it is adequate to simply append the unitary rotations at rate $\Omega$ to the measurement dynamics derived via the CA model, as in \eqref{eq:ZenoMasterEquation}.
\item In the opposite limit $\Omega \gg \kappa$, however, the measurement strength will be severely inhibited.
This can occur for two reasons: (first) The authors of Ref.~\cite{Szombati2020} show that the cavity has no chance of acting as a pointer for qudit dynamics for fast drive $\Omega \gtrsim \Delta \chi^2/\kappa$.
In other words, for any finite $\chi$ and $\kappa$, sufficiently fast rotations $\Omega$ will completely inhibit measurement because the cavity pointer states lack the agility to resolve the qudit dynamics. We shall refer to this below as the ``decoupled regime''.\footnote{Szombati et al.~\cite{Szombati2020} term this the ``quantum rifling'' regime. We understand their result, based around driving the system very fast so that the field only responds to an average state rather than following the dynamics, as a time--continuous analog of bang--bang dynamical decoupling \cite{Viola_1998, Viola_2004_JOMO}.}
(second) One may alternatively reach $\Omega \gg \kappa$ by making the cavity emission very slow; this tends to inhibit measurement as well, simply because the limit of vanishing $\kappa$ is the limit of the system becoming closed to the outside world.  
\end{enumerate}

In summary, our ability to measure the qudit via dispersive coupling may be severely impacted by the relationship between different timescales, including the gate time (recall that $T_G = 2\pi/\Omega$). 
Clearly, the gate time cannot be made arbitrarily short (we must avoid the decoupled regime). On the other hand, our requirement that $\kappa$ be small enough to resolve only a particular qudit transition forces a choice between two options for practical implementations of the Zeno gate.

i) \emph{A Slow Zeno Gate}: We may consider quasi-Markovian dynamics (i.e.~we retain the CA model, and use the steady state solutions \eqref{cavss}, so that we may proceed with analytic expressions). 
The drawback of such an analysis is that it is only valid when the Zeno gate is performed very slowly, due to the need to maintain $\Omega \ll \kappa \lesssim \Delta \chi$. 

ii) \emph{Speeding up the Zeno Gate}: If we shorten the gate time, we must accept that $\Omega \sim \kappa$ will bring some non--Markovian effects into play and erode the validity of the CA measurement model (i.e.,~the cavity will not be describable by coherent states and will retain a memory of the qudit dynamics such that the effective measurement strength at any given time may depend on the history of the system over timescales of order $\kappa^{-1}$). This means we are forced into a transient regime between the extremes of the Markovian and decoupled regimes \cite{koolstra2021monitoring}. 

We briefly discuss the first approach (a slow gate) in some detail, and then move onto full simulation (FS) of the Zeno gate (without the assumption \eqref{eq:rhoPolaron}) in order to better understand the viability of a faster gate.

\subsubsection{A Slow Markovian Zeno Gate: $\Delta \chi \gg \kappa$}

\begin{figure}
\includegraphics[width = .95\columnwidth]{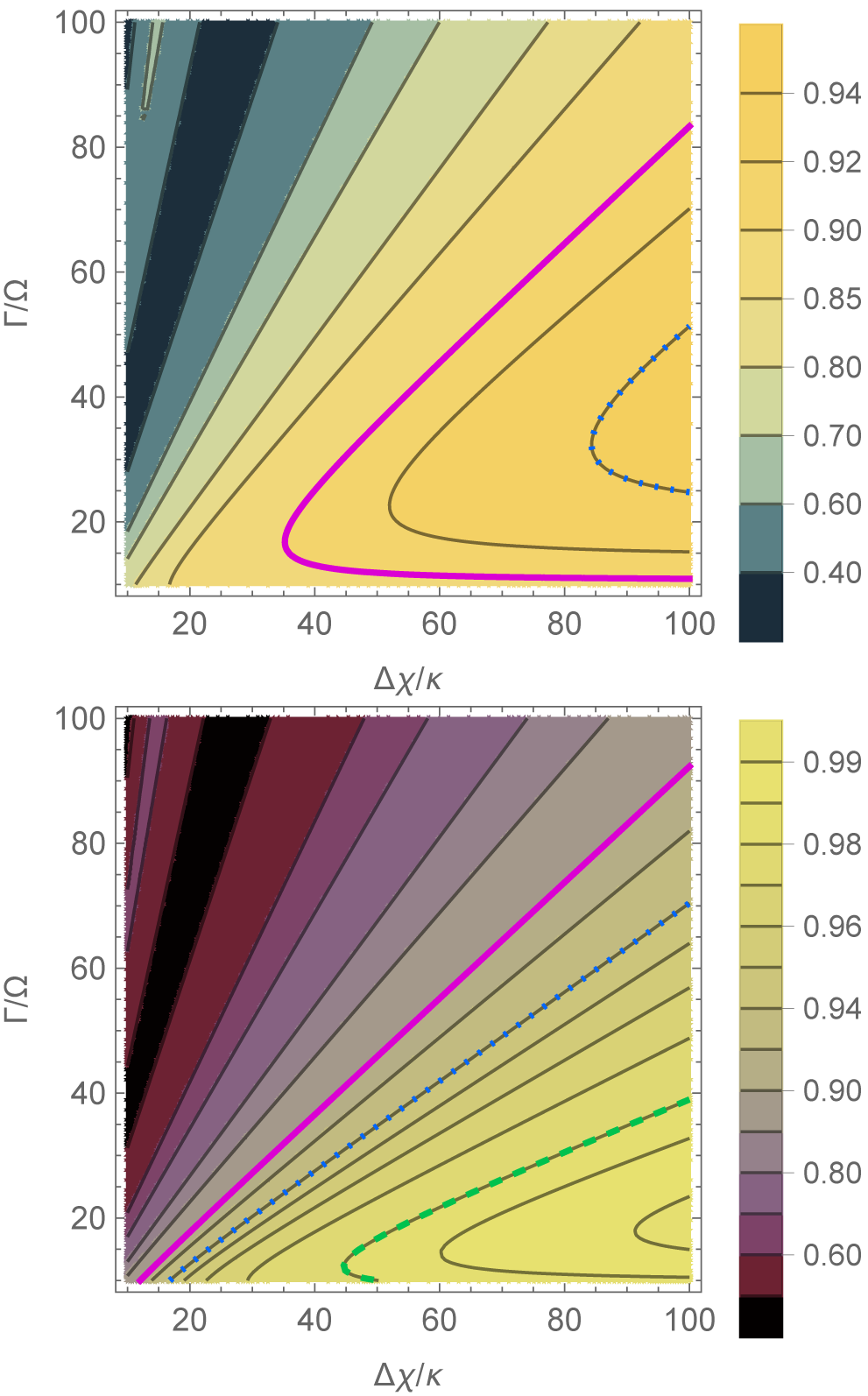}
\caption{Average Zeno gate fidelity in the case of a Markovian cavity response for slow cavity decay $\kappa$, with a slower Rabi drive $\Omega \ll \kappa$, shown as a function of $\Delta\chi/\kappa$ and $\Gamma/\Omega$ [see \eqref{eq:FSecondOrder}]. 
The upper panel shows unheralded gate fidelity while the lower panel shows the improved heralded gate fidelity. 
Specific contours corresponding to certain fidelity values are marked as follows: $\bar{F} = 0.9$ (solid magenta), $\bar{F} = 0.94$ (dotted blue), and $\bar{F} = 0.98$ (dashed green). 
These two-dimensional plots are computed using approximations for the slow Markovian regime defined by $\Omega \ll \kappa \ll \Delta\chi$, with dispersive shifts spaced equally by $\Delta\chi$ (and proportional to qudit excitation number). 
The oscillations in the upper left of each panel are entirely due to resonator--induced phase effects. 
The infidelity $1-\bar{F}_\chi$ from dispersive effects within the two--qubit subspace $\mathcal{S}_Z$ is combined with $\bar{F}_\mathrm{Herald}$ and $\bar{F}_\mathrm{finite~\Gamma}$ as in \eqref{eq:FSecondOrder}. 
Further technical details appear in Appendices \ref{sec:CalcDetails}--\ref{sec:Haar} and \ref{app-Markov}--\ref{app-largechi}, leading to \eqref{eq-FID-fig5}. 
While RIP effects within $\mathcal{S}_Z$ are included in this analysis, RIP effects outside of $\mathcal{S}_Z$ are omitted.
}
\label{fig:2DAverageFid}
\end{figure}

It is possible to make analytic statements in the regime $\Omega \ll \kappa \ll \Delta \chi$, which admits relatively high Zeno gate fidelities. 
We detail the requisite calculations in Appendicies~\ref{app-Markov}--\ref{app-largechi}, and summarize the main findings here, with reference to Fig.~\ref{fig:2DAverageFid}.
The appeals of this regime include 1) an analytic and Markovian description of the qudit measurement, with 2) potentially very well--isolated measurement of transitions to $\ket{fe}$ only.
The measurement rates can be approximated with simple expressions (see Appendix \ref{app-Markov}) and the essential features of the measurement dynamics evaluated analytically (see Appendix \ref{app-largechi}) and then used to evaluate the gate fidelity (see Appendix \ref{sec:Haar}).
We reiterate that this is however at the cost of a very slow gate.
For example, we may consult Fig.~\ref{fig:2DAverageFid} and find that a \emph{heralded} gate fidelity $\bar{F} \gtrsim 0.9$ in this regime requires $\Delta\chi/\kappa \gtrsim 20$ and $\Gamma/\Omega \gtrsim 20$. For realistic choices of other parameters, this implies a gate time on the order of $T_G \gtrsim 20\,\mathrm{\mu s}$.

\subsubsection{Towards a Fast non-Markovian Zeno Gate}

\begin{figure}
    \centering
    \includegraphics[width = \columnwidth]{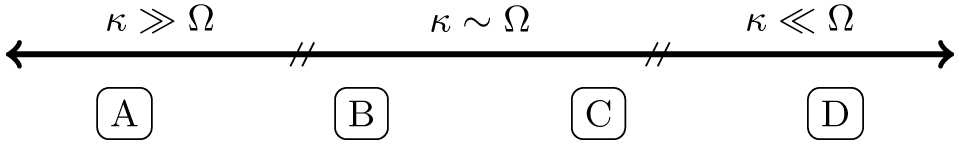}
    \caption{
    Schematic showing the approximate division into different dynamic regimes for the Zeno gate. 
    At the extreme left is the Markovian regime (A) which is characterized by ideal measurement dynamics, and at the extreme right 
    (D) measurement fails entirely. If (D) is reached by making $\kappa$ very small, measurement fails because the system is becoming closed, whereas if it is reached by making $\Omega$ very fast we enter the ``quantum rifling'' $\Omega \gtrsim \Delta \chi^2/\kappa$ \cite{Szombati2020} regime in which the driven qudit transitions effectively decouple from the readout cavity.  
    In between lies a large transient regime that is neither ideal for measurement, nor prohibits dispersively--mediated measurement or dissipation of qudit information. 
    In this work we split this intermediate regime into two for the purposes of discussion. We consider a ``shallow non-Markovian'' (shallow NM) regime (B) defined by $\kappa/\Omega \gtrsim 1$, and a ``deep non-Markovian'' (deep NM) regime (C) defined by $1 \gtrsim \kappa/\Omega$. These divisions are meant to be suggestive rather than exact and may not necessarily apply or be adequate to characterize other problems outside the scope of the present manuscript.
    }
    \label{fig_NM-Regimes}
\end{figure}

Our primary objective in the next section is to investigate Zeno gate performance in non-Markovian (NM) readout regimes, which of practical importance for a faster Zeno gate. 
A few works have considered dispersive readout in such intermediate regimes $\kappa \sim \Omega$ in detail: 
Ref.~\cite{gambetta2008trajectory} details the basis of numerical approaches, which are tractable but lack the analytic clarity of the simpler limiting cases above.   
Experimental work seeking an understanding of the dynamics in the intermediate parameter regime has recently been carried out by Koolstra et al.~\cite{koolstra2021monitoring}. 
The authors of the latter work find that measurement models derived for the Markovian regime lose accuracy for $\Omega\gtrsim \tfrac{1}{2}\kappa$. 
We let this bound orient our discussion going forward. 
In particular, we should expect that any attempts to implement a Zeno gate on a timescale shorter than $4\pi/\kappa$ may lead to a modified measurement strength compared with the predictions of the CA model, due to deviations from the ansatz \eqref{eq:rhoPolaron}. 
For the remainder of this paper we shall qualitatively divide the non-Markovian regime $\kappa \sim \Omega$ into the ``shallow non-Markovian regime'' (shallow NM regime, with $\Omega \lesssim \kappa$), and the ``deep non-Markovian regime'' (deep NM regime, with $\Omega \gtrsim \kappa$) for purposes of discussion (see Fig.~\ref{fig_NM-Regimes}).

\section{Numerics: Practical Implementation of the Zeno Gate in Circuit QED \label{sec_Numerics}}

We here employ two approaches in our numerical explorations.
First, we perform full simulation of the dispersive--coupled qubit and cavity, with cavity decay into an external line (i.e.,~we simulate the dynamics under \eqref{eq_HDisp} and \eqref{eq:OpenCavME}). 
It is also useful however to investigate the dynamics of the Zeno gate using a na\"{i}ve application of the CA model. 
The latter entails computing the equations of motion for the measurement alone (using the coherent state ansatz \eqref{eq:rhoPolaron}), and then simply adding $i[\hat{\rho},\tfrac{1}{2}\boldsymbol{\Omega}\cdot\hat{\boldsymbol{\sigma}}]$ to the equations of motion after the fact. 
We stress that this latter approach is not formally justifiable, and should be treated as an ad hoc approximation of limited validity \cite{Korotkov2016, koolstra2021monitoring}. However, by comparing this na\"{i}ve approach to more exact simulations, we are able to better appreciate how and where the assumptions underlying the CA model break down as we increase $\Omega/\kappa$. 
In practice, the schematic division of Fig.~\ref{fig_NM-Regimes} corresponds approximately to where the na\"{i}ve model reproduces qualitatively the correct dynamics (in the shallow NM regime), versus where it ceases to do so (in the deep NM regime).

In this section we shall consider two figures of merit for the performance of the Zeno gate.
One is the fraction of the population initially in $\ket{ee}$ that is \emph{prevented} from moving into $\ket{fe}$ halfway through the gate. 
Recall that after a duration $t = T_G/2 = \pi/\Omega$, with no measurement \emph{all} of the population initially in $\ket{ee}$ would rotate to $\ket{fe}$.
We always choose the initial qudit state $\tfrac{1}{2}\left(\ket{ee} + \ket{eg} + \ket{ge}+ \ket{gg} \right)$ below, such that this figure of merit reads
\be \label{eq_Xife}
\Xi^{(fe)} = 1 - 4 \bra{fe} \hat{\rho} \ket{fe},
\ee
where the factor of $4$ derives from the $1/4$ probability to be initially in $\ket{ee}$. 
While $\Xi^{(fe)}$ does not directly reveal our effective measurement strength, we will use it as a proxy for the effectiveness of our Zeno blocking; it takes on a value of $0$ for no measurement, and a value of $1$ when population transfer $\ket{ee}\rightarrow\ket{fe}$ is completely inhibited. 
The other figure of merit we consider is the concurrence $\mathcal{C}$ \cite{Wootters1998} in the two--qubit computational subspace $\lbrace \ket{ee},\ket{eg},\ket{ge},\ket{gg} \rbrace$ at the end of the Zeno gate $t = T_G$.   
A value of $\mathcal{C} = 1$ indicates both that all population has returned to the computational subspace, \emph{and} that perfect two--qubit entanglement has been created in that subspace, consistent with perfect implementation of the gate on the chosen initial state. 
Any discrepancies between the evolution and desired gate would then be correctable after the fact by implementation of \emph{local} unitaries. $\mathcal{C} = 1$ indicates primarily that the measurement has done all of the \emph{non-local} evolution involved in the gate, and secondarily that population at the end of the gate time has returned to the intended subspace.

\subsection{Ideal Dispersive Readout Scenario for the Zeno Gate}

\begin{figure*}
\includegraphics[width = \textwidth]{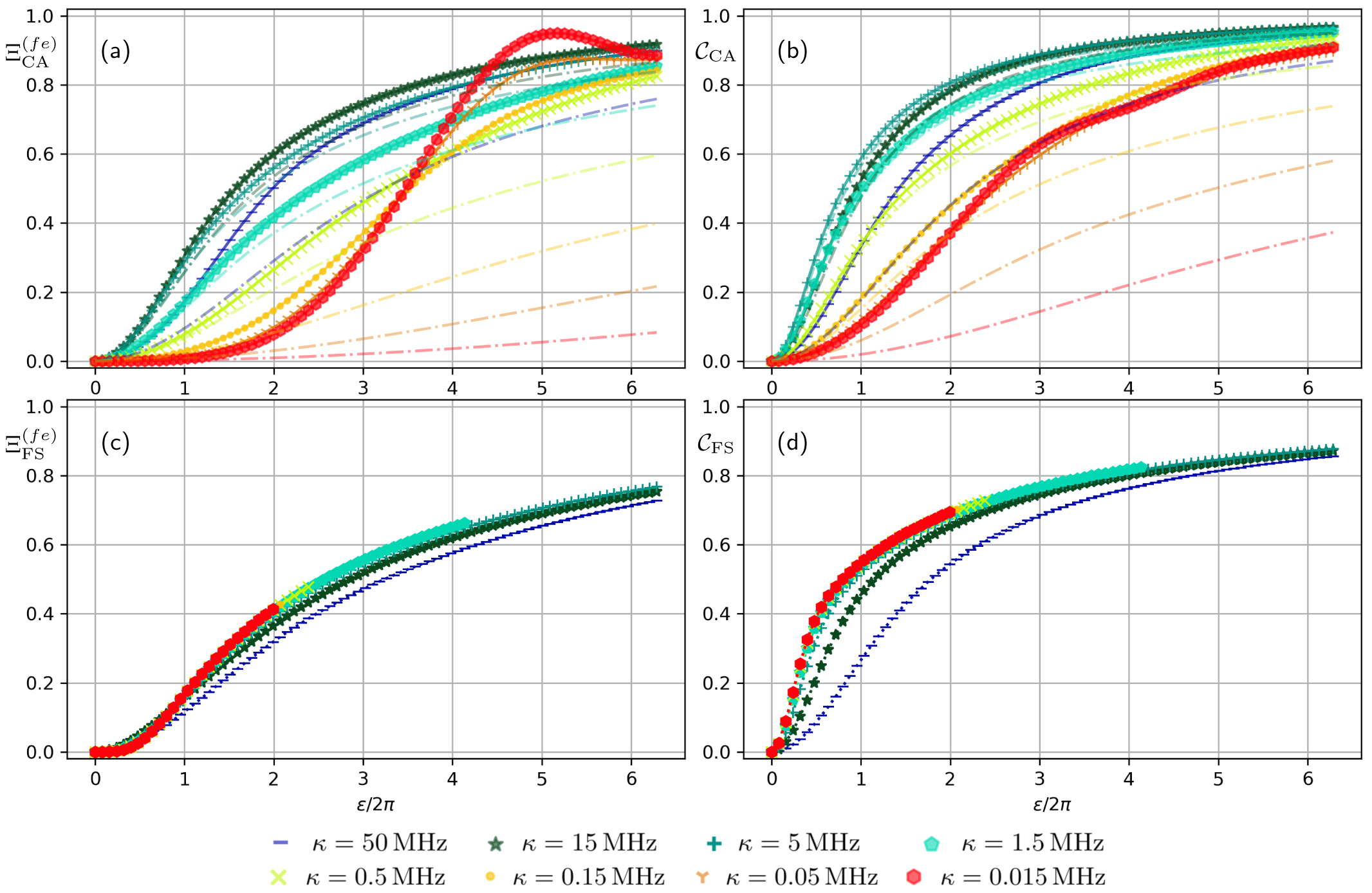}
\caption{
Figures of merit for performance of the Zeno gate under ideal implementation with dispersive measurement. The subspace retention $\Xi^{(fe)}$ \eqref{eq_Xife} (left panels) and concurrence $\mathcal{C}$ (right panels), are shown as a function of the measurement tone driving amplitude $\varepsilon$, for the Zeno gate implemented over a time $T = 1\,\mathrm{\mu s}$ (i.e.,~for $\Omega/2\pi = 1\,\mathrm{MHz}$). We use the idealized set of system parameters $\chi_{fe}/2\pi = 15\,\mathrm{MHz}$ with all other $\chi = 0$. 
The initial state is $\tfrac{1}{2}(\ket{e} + \ket{g})\otimes(\ket{e}+\ket{g})\otimes \ket{0}$. 
Results are shown for $\kappa$ values ranging from the Markovian regime to the deep NM regime. Curves linking the solid symbols obtained from na\"{i}ve application of the CA model (described in Sec.~\ref{sec-CA-model}) are plotted in panels (a) and (b), while corresponding curves obtained from full simulation of \eqref{eq_HDisp} and \eqref{eq:OpenCavME} are shown in panels (c) and (d). 
The low--opacity dash-dotted curves in panels (a) and (b) show idealized dynamics with phase rotations artificially suppressed (i.e.,~$\Upsilon =0$) (see text for discussion). 
Agreement between the theory (top) and simulated (bottom) results is the best in the shallow NM regime, i.e.~for the curves plotted with $\kappa \gtrsim 1.5\,\mathrm{MHz}$.
All simulations (c,d) are made via the unconditional master equation (describing an unheralded Zeno gate) using a cavity Fock space truncated at $N \leq 50$ photons.}
\label{fig-ideal-chi}
\end{figure*}

We first model dispersive measurement with a very optimistic set of parameters to provide an idealized reference point. 
Suppose that we are able to choose $\chi_{fe}/2\pi = 15\,\mathrm{MHz}$, while all other $\chi$ values are zero (which eliminates both RIP effects and decoherence within the computational subspace $\mathcal{S}_Z$).
Such a choice of parameters is clearly at odds with the actual scaling of dispersive shifts with qudit excitation number. 
Simulations for this idealized system with dispersive readout are shown in Fig.~\ref{fig-ideal-chi}, with $\Omega/\kappa$ spanning roughly $0.1$ to $500$. 

Despite the un-realistic idealization of the dispersive shifts implicit in this first set of simulations, a number of important qualitative features are revealed by Fig.~\ref{fig-ideal-chi}. 
We plot the results of a na\"{i}ve application of the CA model in the upper panels (a,b): Solid curves with thick markers denote the results of the full CA model, while lower--opacity dash--dotted curves show the corresponding analysis with $\Upsilon$ artificially set to zero (i.e.~the dash--dotted curves illustrate dynamics due to a measurement without phase resonator--induced effects).
Thus, the difference between the solid and dash--dotted curves allow us to determine the extent to which RIP and AC--Stark effects are playing a role in our gate. 
It is clear even upon superficial inspection that these effects play a large role at higher drive amplitudes $\varepsilon$, and deeper in the NM regime: This is ostensibly due to the cavity collecting a higher average photon number, which exaggerates RIP contributions $\varepsilon\,\Im[\alpha_\ell - \alpha_j]$. 
The results of full simulations (FS) appear in the lower panels (c,d) of Fig.~\ref{fig-ideal-chi}. 
These FS results are the most accurate form of analysis we perform, and the discrepancies between these curves and those from the na\"{i}ve CA model above allow us to understand the role of NM readout effects on the Zeno gate. 
These reveal that the CA model is qualitatively correct in the shallow NM regime, but that its predictions become increasingly poor as we scale into the deep NM regime.

\begin{figure*}
\includegraphics[width = \textwidth]{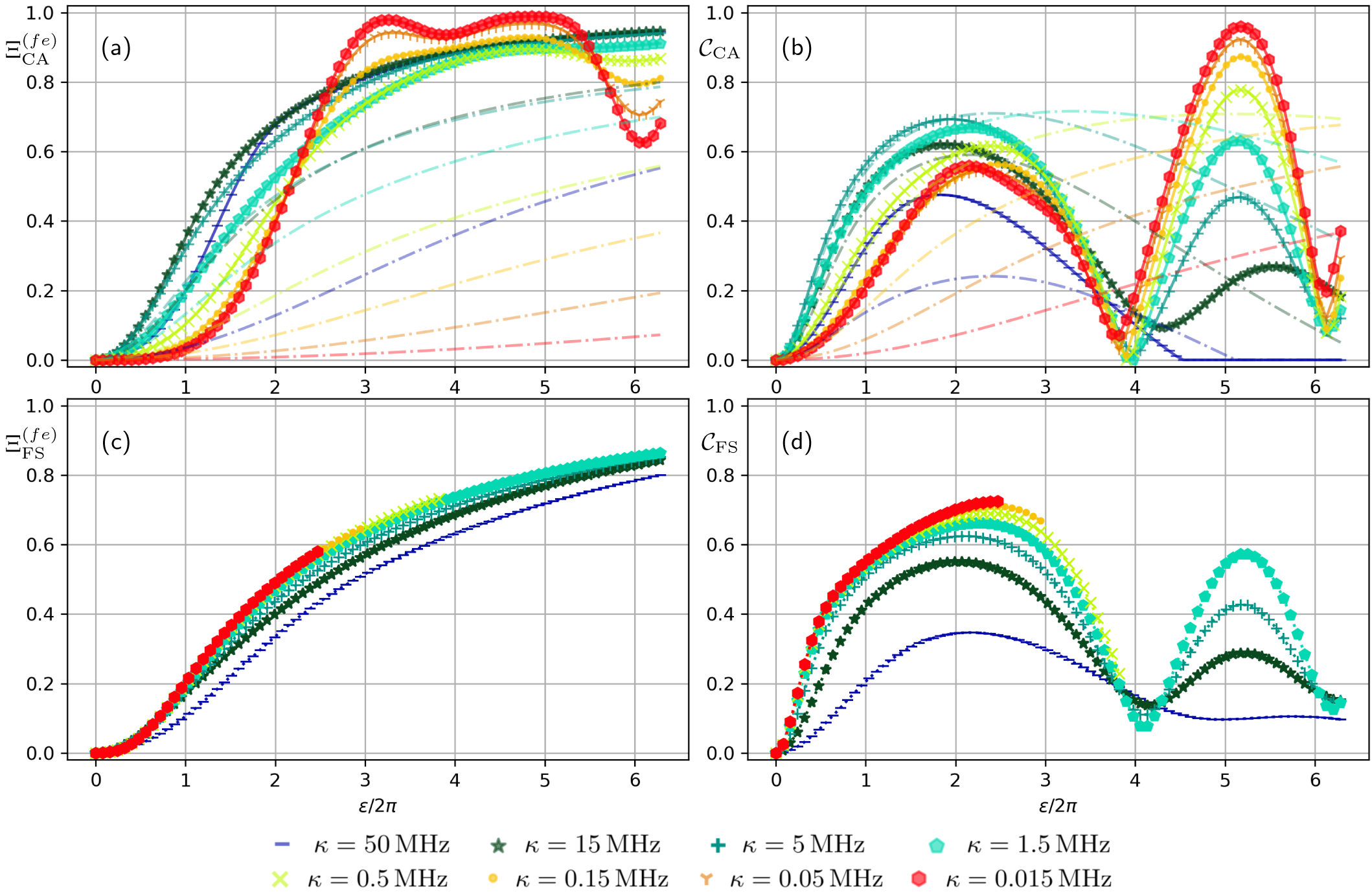}
\caption{
Figures of merit for performance of the Zeno gate under realistic implementation with dispersive measurement for short gate time, $T = 1\,\mathrm{\mu s}$ (i.e.,~for $\Omega/2\pi = 1\,\mathrm{MHz}$). The subspace retention $\Xi^{(fe)}$ \eqref{eq_Xife} (left panels) and concurrence $\mathcal{C}$ (right panels), are shown as a function of the measurement tone driving amplitude $\varepsilon$.
We use dispersive shift values $\chi_{fe}/2\pi = 15\,\mathrm{MHz}$, $\chi_{fg}/2\pi =  \chi_{ee}/2\pi = 9\,\mathrm{MHz}$, $\chi_{eg}/2\pi = \chi_{ge}/2\pi = 4\,\mathrm{MHz}$, and $\chi_{gg} = 0$. 
The initial state is $\tfrac{1}{2}(\ket{e} + \ket{g})\otimes(\ket{e}+\ket{g})\otimes \ket{0}$. 
Results are shown for $\kappa$ values ranging from the Markovian regime to the deep NM regime. Curves linking the solid symbols obtained from na\"{i}ve application of the CA model (described in Sec.~\ref{sec-CA-model}) are plotted in panels (a) and (b), while corresponding curves obtained from full simulation of \eqref{eq_HDisp} and \eqref{eq:OpenCavME} are shown in panels (c) and (d). 
The low--opacity dash-dotted curves in panels (a) and (b) show idealized dynamics with phase rotations artificially suppressed (i.e.,~$\Upsilon =0$). 
All simulations (c,d) are made with the unconditional master equation for an unheralded Zeno gate and use a cavity Fock space truncated at $N \leq 80$ photons.
} \label{fig-res-chi}
\end{figure*}

\subsection{Realistic Zeno Gate Implementation with Dispersive Readout}

\begin{figure*}
\includegraphics[width = \textwidth]{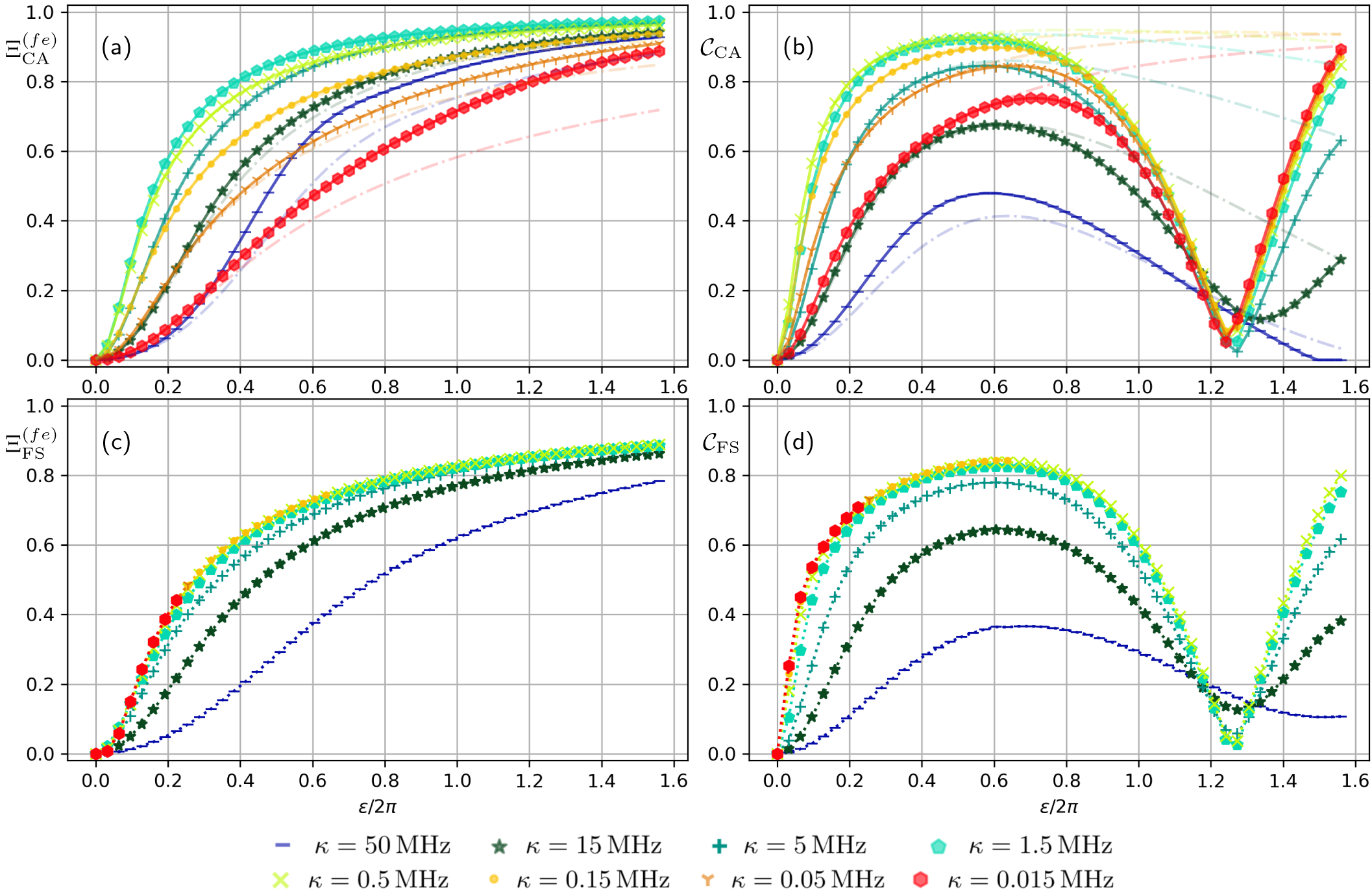}
\caption{
Figures of merit for performance of the Zeno gate under realistic implementation with dispersive measurement for a longer gate time, $T = 10\,\mathrm{\mu s}$ (i.e.,~for $\Omega/2\pi = 0.1\,\mathrm{MHz}$). The subspace retention $\Xi^{(fe)}$ \eqref{eq_Xife} (left panels) and concurrence $\mathcal{C}$ (right panels), are shown as a function of the measurement tone driving amplitude $\varepsilon$. 
We use dispersive shift values $\chi_{fe}/2\pi = 15\,\mathrm{MHz}$, $\chi_{fg}/2\pi =  \chi_{ee}/2\pi = 9\,\mathrm{MHz}$, $\chi_{eg}/2\pi = \chi_{ge}/2\pi = 4\,\mathrm{MHz}$, and $\chi_{gg} = 0$. 
The initial state is $\tfrac{1}{2}(\ket{e} + \ket{g})\otimes(\ket{e}+\ket{g})\otimes \ket{0}$. 
Results are shown for $\kappa$ values ranging from the Markovian regime to the deep NM regime. Curves linking the solid symbols obtained from na\"{i}ve application of the CA model (described in Sec.~\ref{sec-CA-model}) are plotted in panels (a) and (b), while corresponding curves obtained from full simulation of \eqref{eq_HDisp} and \eqref{eq:OpenCavME} are shown in panels (c) and (d). 
The low--opacity dash-dotted curves in panels (a) and (b) show idealized dynamics with phase rotations artificially suppressed (i.e.,~$\Upsilon =0$). 
All simulations (c,d) are made via the unconditional master equation, and the points plotted have converged within a cavity Fock space truncated at $N \leq 60$ photons.}
\label{fig-res-chi-slow}
\end{figure*}

We now present results of  numerical simulations for a dispersive Zeno gate with realistic system parameters.
Figs.~\ref{fig-res-chi} and \ref{fig-res-chi-slow} show representative plots for two sets of parameters, one for a faster gate time $T = 1\,\mathrm{\mu s}$ (Fig.~\ref{fig-res-chi}) and one for a slower gate time $T = 10\,\mathrm{\mu s}$ (Fig.~\ref{fig-res-chi-slow}).
The system parameters are in both cases specified as
$\chi_{gg} = 0$, $\chi_{eg}/2\pi = \chi_{ge}/2\pi = 4\,\mathrm{MHz}$, $\chi_{fg}/2\pi = \chi_{ee}/2\pi = 9\,\mathrm{MHz}$, and $\chi_{fe}/2\pi = 15\,\mathrm{MHz}$.
Note that these values are still slightly idealized compared with those shown in Fig.~\ref{fig-linestrength}.
However they share the most essential feature of a large dispersive gap $\Delta \chi/\kappa$. 

All of Figs.~\ref{fig-ideal-chi} through \ref{fig-res-chi-slow} may be read individually in much the same way we have just analyzed Fig.~\ref{fig-ideal-chi}: Comparison of curves within the upper CA panels shows the impact of RIP effects, while comparison of the upper and lower panels illustrates the impact of NM effects. 
We may furthermore compare the three figures to each other, and draw some conclusions about the dispersive Zeno gate as a whole:

1) Na\"{i}ve application of the CA model, in which coherent states are assumed and the non--commuting unitary dynamics \eqref{HRabi_3x2} are appended in an ad hoc manner, remains qualitatively correct throughout the shallow NM regime. This is consistent with the observations of \cite{koolstra2021monitoring}. 

2) RIP effects, evaluated by comparing the solid and dash--dotted curves within the upper (CA) panels, are significant in all of the figures. Fig.~\ref{fig-ideal-chi} differs from Figs.~\ref{fig-res-chi} and \ref{fig-res-chi-slow} in that the latter (realistically) include RIP effects \emph{within} the computational subspace. Comparison of the faster gate (Fig.~\ref{fig-res-chi}) against the slower (Fig.~\ref{fig-res-chi-slow}) reveals that the impact of RIP effects are lessened for a slow gate using lower drive amplitude $\varepsilon$. 
Looking across all three of our figures, we find that RIP effects play at least as large a role in the dynamics as NM effects, for many of the parameter ranges we consider. 

3) Simulations reveal that decent gate performance is possible deeper into the NM regime than we might have expected: Both the CA model, and the limiting case of the decoupled regime \cite{Szombati2020}, suggest that the deep NM regime should be quite poor for the Zeno gate. 

However our simulations do not bear out the analytic prediction that the measurement ring--up time will be prohibitively long for very small $\kappa$ values. 
In Appendix \ref{app-moresim} we speculate that this may be explained by the Rabi drive effectively strengthening the measurement and thereby speeding up of the cavity response in certain regimes, before the Rabi rotations are fast enough to push us into the decoupled regime.

4) While it is absolutely necessary to leave the Markovian regime and to enter the shallow NM regime to operate the Zeno gate for the realistic parameters of Figs.~\ref{fig-res-chi} and \ref{fig-res-chi-slow} over modest gate times, the gains obtained by continuing into the deep NM regime are only marginal (despite the unexpected benefits described in point 3).

5) Fig.~\ref{fig-res-chi-slow}, with the longer gate time, shows overall better performance than Fig.~\ref{fig-res-chi}. 
We find higher concurrence generation with lower drive powers compared with Fig.~\ref{fig-res-chi}, which overall constitute a significant improvement in the Zeno gate. 
Thus a high fidelity Zeno gate can be achieved with realistically non-ideal parameters, but our best results are still attained by running the gate slowly. The slowness of the gate will, in practice, be limited by other coherence times of the system that are not considered in the present simulations. 
It is also evident that a longer (slower) gate requires a smaller drive $\varepsilon$ compared to a fast gate (compare the range of $\varepsilon$ in Fig.~\ref{fig-res-chi} against Fig.~\ref{fig-res-chi-slow}), suggesting that the photon ring--up time may be a significant contributor to the ineffectiveness of fast gates.

6) The higher fidelity gates shown throughout these figures rely on values of $\varepsilon$ that could result in large cavity photon numbers (if $\ket{fe}$ is populated) for smaller values of $\kappa$. 
Large photon numbers can break the dispersive approximation, and study of measurement in such regimes is an area of contemporary interest in the literature \cite{Sank_2016, Walter_2017, Lescanne_2019, Malekakhlagh_2020, Petrescu_2020, Gusenkova_2021, Shillito_2022}. 
While analysis of the Zeno gate based on transverse coupling, beyond the dispersive approximation, may be a fruitful direction for future work, it remains beyond the scope of the present manuscript. 

In summary, our analysis for the Zeno gate under dispersive readout predicts that this implementation of the $N=2$ Zeno gate can be operated at reasonably high fidelity (i.e., achieving concurrence $\mathcal{C}\gtrsim 0.8$), using dispersive measurement and a realistic set of transmon parameters, with a moderate gate duration characterized by $T_G \gtrsim \kappa^{-1} \gtrsim \Delta \chi^{-1}$, as opposed to the more ideal separation of timescales $T_G \gg \kappa^{-1} \gg \Delta \chi^{-1}$.
Moderate measurement tone amplitudes are adequate, and a slower gate is desirable to the extent that is allowed by the transmon coherence times. 
Finally, we emphasize that all the results in this section were developed using the unconditional (unheralded) system evolution.  This should be understood as a baseline from which we can only improve by making use of the finite--efficiency detection for heralding, as discussed in Sec.~\ref{sec:constraints}.

\section{Zeno Gate via Longitudinal readout \label{sec-longitude}}

\begin{figure*}
\includegraphics[width = \textwidth]{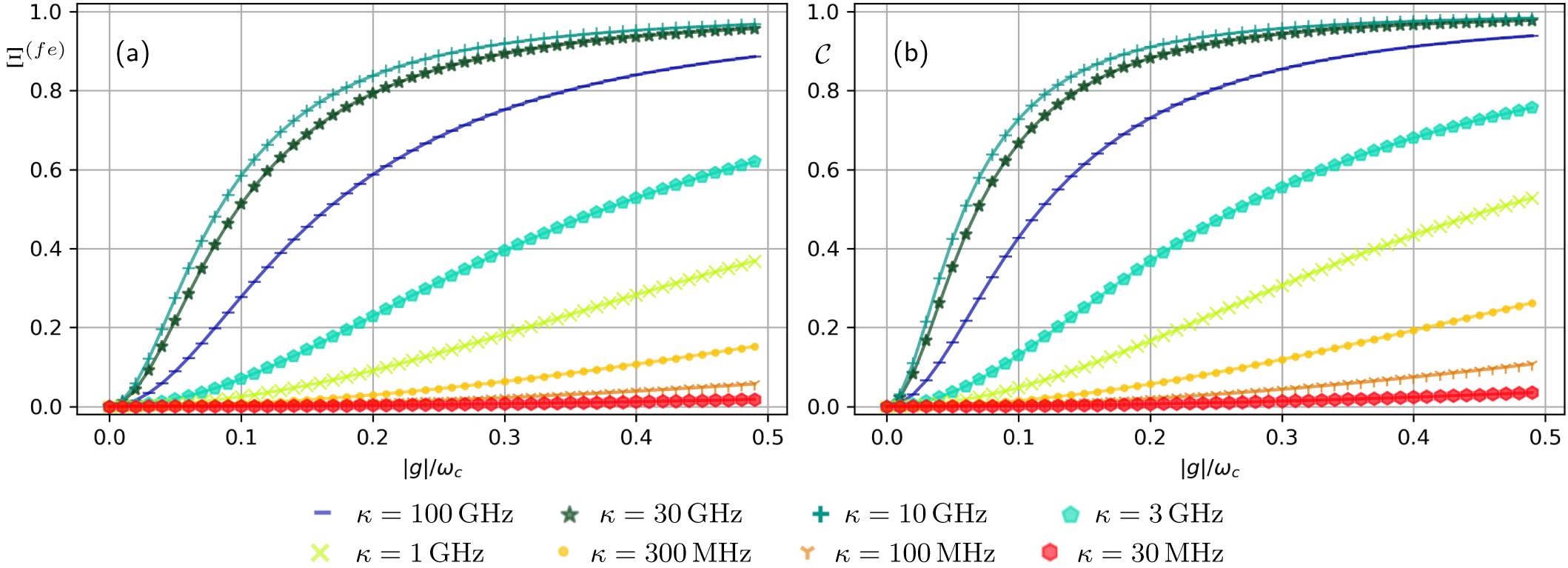}
\caption{
Figures of merit for performance of the Zeno gate under ideal implementation with longitudinal measurement. The subspace retention $\Xi^{(fe)}$ \eqref{eq_Xife} (left panels) and concurrence $\mathcal{C}$ (right panels), are shown as a function of the measurement tone driving amplitude $\varepsilon$, for the Zeno gate implemented over a time $T_G = 100\,\mathrm{ns}$ (i.e.,~for $\Omega/2\pi = 10\,\mathrm{MHz}$). 
The initial state is $\tfrac{1}{2}(\ket{e} + \ket{g})\otimes(\ket{e}+\ket{g})\otimes \ket{0}$. 
The cavity resonance is set at $\omega_c/2\pi = 1\,\mathrm{GHz}$ and the system-cavity coupling terms are $g_{fe} = |g|$ and $g_j = -|g|$ for $j \neq fe$, which eliminates both decoherence in the computational subspace $\mathcal{S}_Z$ and any relative phase dynamics (see Appendix~\ref{app-measurement}). 
} 
\label{fig-ideal-lon}
\end{figure*}

We briefly explore here an alternative to dispersive measurement that could be investigated in depth for future realizations of the Zeno gate. 
There has been considerable interest in engineering ``longitudinal'' couplings of the form
\be 
\hat{H}_\mathrm{lon} = \omega_c\,\hat{a}^\dag\hat{a}+\sum_j \left\lbrace\omega_j + g_j(\hat{a}^\dag + \hat{a})  \right\rbrace\ket{j}\bra{j}
\ee
for readout, in place of the transverse coupling \eqref{eq_Htransverse} (see \cite{blais2020circuit} for an overview). 
This coupling also leads to a description of readout in terms of coherent pointer states with a measurement strength \eqref{meas-strength}, with pointer state distance now entirely determined by the relative values of $g_j$ for different states. (For further details, see Appendix \ref{app-measurement}.)
This implies that if the $g_j$ can be precisely engineered independent of one another in practice, then longitudinal readout offers the possibility of realizing dynamics closer to those of \eqref{eq:ZenoMasterEquation}.
Examples of longitudinal readout design include more elaborate circuits in the spirit of \cite{Didier_2015_Longitudinal}, and sideband--based single--quadrature measurements \cite{MartinLeighS_2020Thesis, hacohen2016noncommuting,Eddins_2018}.
If an intrinsically--degenerate measurement can be realized by these or other methods, it would mitigate decoherence of the computational subspace $\mathcal{S}_Z$ \emph{without} constraining $\kappa$ as was necessary in the dispersive case above.

Fig.~\ref{fig-ideal-lon} summarizes the performance of the longitudinal Zeno gate idealized in the same spirit as the dispersive analysis of Fig.~\ref{fig-ideal-chi}. 
The ideal longitudinal readout conditions maximize the coupling gap $\Delta g_{fe,j}$, while minimizing the coupling gap $\Delta g_{j,\ell}$ for $j,\ell \neq fe$. 
Under these idealized conditions we see that the Zeno gate performance is much cleaner than even the idealized version of dispersive measurement in Fig.~\ref{fig-ideal-chi}.
The curves in panels (a) and (b) of Fig.~\ref{fig-ideal-chi} provide a sense of the parameter ranges that would be required for implementation of a longitudinal Zeno gate. 
The coupling parameters are chosen such that $g_{fe} = |g|$ and $g_j = -|g|$ for $j \neq fe$, so as to eliminate both decoherence in the computational subspace $\mathcal{S}_Z$ and the undesirable relative phase dynamics, as discussed in detail in Appendix~\ref{app-measurement}. 
The longitudinal Zeno gate implemented with this parameter set is relatively rapid, with a gate time $T_G = 100\,\mathrm{ns}$ which is partly responsible for the relatively demanding parameters quoted in Fig.~\ref{fig-ideal-lon}. However we note that the cavity decay rates $\kappa$ for which this gate is expected to perform well are much higher than the corresponding values under dispersive readout (Fig.~\ref{fig-ideal-chi}). 
Indeed, they are so rapid that issues of non-Markovianity and cavity memory are completely irrelevant (with $\kappa$ on the order of $\mathrm{GHz}$, the gate time would need to be on the order of $\mathrm{ns}$, rather that $\mathrm{\mu s}$ to create any such issue).

In short, longitudinal readout may offer a route to escape some of the punishing timescale tradeoffs that are fundamental to realizing the Zeno gate with dispersive readout (\emph{if} the $g_j$ can be engineered degenerately so as to directly define $\mathcal{S}_Z$, while providing a strong Zeno block against leakage out of $\mathcal{S}_Z$). 
We therefore suggest that further development of this and other readout methods may enable improved Zeno gate performance. We also point out that Fig.~\ref{fig-ideal-lon} may be viewed as ideal behavior for a Zeno gate, and used as a reference against which to compare the qualitative behavior in the dispersive figures preceding it.

\section{Discussion \& Conclusions}
\label{sec:Conclusion}

We have derived a multi-qubit entangling gate based on the Zeno effect and studied its performance under a variety of models and assumptions, both idealized and realistic.
The gate can ideally entangle an arbitrary number of qubits, and effectively implements a \textsc{cphase} or \textsc{n--cphase} gate. 
The operation of the Zeno gate is furthermore locally equivalent to a \textsc{cnot} or Toffoli gate. 
Both the $N=2$ and higher order Zeno gates make use of a single local Rabi rotation applied to a qutrit, and then uses the Zeno effect to transform this trivial Hamiltonian into a universal gate that can in principle be used for quantum computation.

We have shown that analysis of generic non-projective measurement required to implement the Zeno gate provides concise, closed--form expressions encapsulating both leakage through the Zeno blocking, and dephasing due to both ideal and non-ideal measurement. 
Our analysis of the effect of finite strength non-ideal Zeno measurements also characterizes the fidelity of a heralded gate, which is found to have greatly improved fidelity over that obtainable without detection.

We have then investigated two specific implementations of the Zeno gate, using measurement protocols grounded in circuit QED. Specifically, we have emphasized detailed numerical analysis for a Zeno gate implemented by dispersive readout, studying both idealized and more realistic system parameters, in different dynamical regimes and for a range of gate times.
We have quantified the impacts of non-Markovian readout effects for realistic parameters in regimes where the cavity memory is non-negligible, illustrating the impact of these effects and their interaction with resonator induced phase effects. Our analysis of dispersive readout assumes that the dispersive approximation remains appropriate: 
High cavity photon numbers, however, can stress the validity of the dispersive approximation, and lead to a wide variety of complex effects that are detrimental to readout, which in practice means that $\varepsilon/\kappa$ cannot be made arbitrarily large. Further discussion pertinent to this point may be found in e.g.~Refs.~\cite{Sank_2016, Walter_2017, Lescanne_2019, Malekakhlagh_2020, Petrescu_2020, Gusenkova_2021, Shillito_2022}.

We expect that several realistic avenues exist for improved gate operation. 
We have demonstrated reasonable fidelities under unconditional Lindbladian dynamics, and argue that these may be regarded as a lower bound upon which one can improve with finite efficiency measurements and post--selection. 
Additional improvements could be considered in future work, including
shaping the measurement tone $\varepsilon(t)$ to minimize the impact of a slow ring--up for the readout resonator, or
shaping the Rabi drive $\Omega(t)$ so that the unitary rotation is slower while the measurement rings up and maintains a steady $\Gamma/\Omega$ throughout the evolution. 
Another option is to better control the phase rotations $\Upsilon$ (containing resonator--induced and AC Stark effects). 
These also impact entanglement generation, and could either be suppressed to create a pure Zeno gate, or accepted and used in a hybrid between the Zeno and RIP \cite{Cross_2015, paik2016experimental} gate.
The former improvements of shaping the measurement and Rabi drives may be accomplished with additional cavity drive(s) \cite{blumenthal2021}, and/or further unitary controls applied to specific qudit transitions. 
Such phase control could be done on average (unconditionally), or better, be implemented as feedback based on finite--efficiency readout, in which case the gate time might also be adjusted based on knowledge of run--specific phase noise.
This is related to a third potential improvement: It should be possible to enhance the \emph{heralded} gate fidelity by making use of quantum feedback \cite{Zhang_2017_feedback} more generally, e.g.,~by varying $\Omega$ according to the measurement signal. 
Such feedback is known to be optimal in similar contexts of measurement-based entanglement generation \cite{martin2017optimal, Martin_2019_FlorEntangle}. 
Recent progress on dispersive readout in the shallow non-Markovian regime \cite{koolstra2021monitoring} required for a moderately fast Zeno gate suggests that such a feedback--assisted Zeno gate might be feasible in practice.
Fourth and last, the most severe restrictions on timescales required for Markovian readout and high gate fidelities that we have discussed for dispersive measurement would be significantly relaxed in an implementation using longitudinal readout instead. 
We have demonstrated that there exist idealized parameter regimes for which longitudinal readout could, in principle, yield a gate that is both faster and more accurate than the dispersive modeling we have performed.

Implementations of the Zeno gate using more qubits and/or the improvements detailed above constitute a promising avenue for future research and development of this method.  
Our analysis of the Zeno gate under dispersive readout in this work but one example. 
In principle one could perform a similar analysis for a broad variety of readout techniques to understand how to bring the Zeno gate to many different experimental platforms. 

Finally, we look beyond the Zeno gate proposed in this work, which uses a qutrit and $N-1$ qubits to perform an \textsc{n--cphase} gate,
to take note of similar concepts in use in other areas of the literature. 
We have already remarked that there is some connection between the dynamics of the Zeno gate and blockade methods \cite{isenhower2010demonstration}. 
There is also a connection between our Zeno gate and the use of dissipation engineering to stabilize cat qubits or GKP qubits \cite{Mirrahimi_2014, Guillaud_2019, Lescanne_2020, Gautier_2022, Sellem_2022, sellem2023stability, gautier2023designing, Regent2023adiabatic}.
There, the stability of e.g.,~the cat codespace within a bosonic mode is enforced by an engineered two--photon dissipation.
In the present manuscript we defined a Zeno gate that would pull us out of the computational subspace $\mathcal{S}_Z$ without dissipation / measurement, but which then becomes a desirable operation when strong dissipative confinement to $\mathcal{S}_Z$ is added. 
Similarly, operations on a dissipative cat qubit rely on slow operations that would push the bosonic cavity mode out of the desired qubit codespace, but using a strong and specifically--engineered dissipation channel stabilizes these operations to the desired subspace. 
Thus, while qubits encoded in bosonic modes are a quite different physical system than the one we consider here, the Zeno effect can play a similar role in those systems as well. 
More broadly, this again suggests that it would be quite interesting to explore Zeno gates across a wider variety of architectures.

\begin{acknowledgments}
We thank Vinay Ramasesh and Machiel Bloch for helpful discussion. 
PL and KBW also acknowledge conversations with Mazyar Mirrahimi, Pierre Rouchon, Alain Sarlette, Fran\c{c}ois--Marie Le R\'{e}gent, J\'{e}r\'{e}mie Guillaud, and Ronan Gautier, that clarified the connection between our Zeno gate and gate implementations on dissipative cat qubits.
This work was supported by Laboratory Directed Research and Development
(LDRD) funding from Lawrence Berkeley National Laboratory, provided by the U.S. Department of Energy, Office of Science under Contract No.~DE-AC02-05CH11231, by Grant No.~2020166 from the United States-Israel Binational Science Foundation (BSF), and by Israel Science Foundation (ISF). 
LM was supported by the National Science Foundation Graduate Fellowship Grant No.~(1106400) and the Berkeley Fellowship for Graduate Study. 
PL and KBW  were partially supported by the U.S. Department of Energy, Office of Science, National Quantum Information Science Research Centers, Quantum Systems Accelerator (QSA).
DB acknowledges support by the Berkeley Center for Quantum Information and by the Australian Research Council (project numbers FT190100106, DP210101367,
CE170100009). 
Publication made possible in part by support from the Berkeley Research Impact Initiative (BRII) sponsored by the UC Berkeley Library.
\end{acknowledgments}

\appendix
\section{Exact and approximate expressions for the effective single-qubit Zeno fidelity}
\label{sec:CalcDetails}
The effective single qubit master equation \erf{eq:QubitZenoME} is a set of linear, first-order differential equations. As $\tilde{\rho}(0)$ is real, symmetric, and of unit trace, and $\tilde{H}$ is purely imaginary, the master equation reduces to a set of two coupled equation in two real variables, which may be solved exactly by diagonalization. The exact solution for $\tilde{\rho}_{11}(t)$ given the initial condition $\tilde{\rho}_{11}(0) = |1\rangle\langle1|$ is 
\begin{align}
\label{eq:AppendixExactRho11}
\tilde{\rho}_{11}(t=2\pi/\Omega) =& \frac{1}{2}\Bigg[1 + e^{-\pi \Gamma/2\Omega}\cosh\left(\frac{\pi A}{2\Omega}\right) \\ \nonumber
&+ \frac{\Gamma}{A}e^{-\pi \Gamma/2\Omega}\sinh\left(\frac{\pi A}{2\Omega}\right)\Bigg] \\ \nonumber
A &\equiv \sqrt{\Gamma^2 - 16 \Omega^2}.
\end{align}
To derive an approximate expression that is much simpler, we use the fact that $\Omega \ll \Gamma$ to expand $A \approx \Gamma - 8\Omega^2/\Gamma$, which yields
\begin{align}
&\tilde{\rho}_{11} \approx \frac{1}{2} \\ \nonumber
&+ \frac{e^{-\pi\Gamma/2\Omega}}{4}\left[e^{\pi \Gamma/2\Omega} e^{-4\pi\Omega/\Gamma} + e^{-\pi \Gamma/2\Omega} e^{4\pi\Omega/\Gamma}\right] \\ \nonumber
&+\frac{\Gamma}{4}\frac{e^{-\pi\Gamma/2\Omega}}{\Gamma-8\Omega^2/\Gamma} \left[e^{\pi \Gamma/2\Omega} e^{-4\pi\Omega/\Gamma} - e^{-\pi \Gamma/2\Omega} e^{4\pi\Omega/\Gamma}\right]
\end{align}
Dropping the $8\Omega^2/\Gamma$ term in the denominator and all terms that are exponentially small in $\Gamma$ yields the expression used in the main text
\begin{align}
\tilde{P}_0 = 1-\tilde{\rho}_{11} \approx \frac{1-e^{-4\pi\Omega/\Gamma}}{2}
\end{align}
The above approximation works very well. Furthermore, the difference between this expression and the exact expression ($1-\tilde{\rho}_{11}$ using \erf{eq:AppendixExactRho11}) is negligible. 

A similar calculation was performed to derive $F_\text{herald}$. There, we used an approximate solution to \erf{eq:HNoJump}. The exact solution is
\begin{align} \label{eq:PsiHerExact}
    \psi_{ee}(t) &= e^{-\pi\Gamma/2\Omega}\left[\cosh\left(\frac{\pi B}{2\Omega}\right) + \frac{\Gamma}{B}\sinh\left(\frac{\pi B}{2\Omega}\right) \right] \\ \nonumber
    B &\equiv \sqrt{\Gamma^2 - 4\Omega^2}
\end{align}
The simplifications leading to \erf{eq:HNoJumpSol} are almost identical to those done above. 

\section{Agreement between Bell state fidelity and average fidelity}
\label{sec:Haar}

To compute average fidelities, one wishes to integrate $\int F(|\psi\rangle)d\psi$ using the Haar measure \cite{ozols2009generate}. Using the fact that the Cartesian product of $n$ Gaussians is spherically symmetric, one can construct a Haar-uniform integral as
\begin{align}
\int F&(\psi_1, \psi_2, ... \psi_d)d\psi \\ \nonumber
=& \frac{1}{\mathcal{N}}\int F \left(\frac{a_1+i b_1}{r}, \frac{a_2+i b_2}{r},... \frac{a_d+i b_d}{r} \right) \\ \nonumber
&\times e^{-r^2} \prod_i da_i db_i \\ \nonumber
r &= \left(\sum_i a_i^2 + b_i^2\right)^{1/2}
\end{align}
where $\mathcal{N}$ is a normalization factor.

We begin with the quantities in Sec.~\ref{sec:FiniteGamma}. The expressions for $F_{\text{finite }\Gamma}^{(1)}$ and $F_\text{herald}$ only depend on $|\psi_{ee}|$, which yields highly symmetric integrals. The Haar measure over a two-qubit Hilbert space has the symmetry of a 7-sphere, which is broken into the symmetry of a 1-sphere (circle) and a 5-sphere by the $|\psi_{ee}|$ dependence of $F$. Thus we first analytically integrate over these spheres using a higher dimensional analog of cylindrical coordinates, leaving nontrivial integrals only over the remaining two radial coordinates
\begin{align}
\mathcal{N} &= \int e^{-r^2} d^8r = \pi^4 \\ \nonumber
m &= \sqrt{a_{ee}^2+b_{ee}^2} \\ \nonumber	
n &= \sqrt{a_{eg}^2+b_{eg}^2 + a_{eg}^2 + a_{eg}^2 + a_{gg}^2+b_{gg}^2} \\ \nonumber
\bar{F} &= \frac{1}{\mathcal{N}} \int F\left(\frac{m}{r}\right) e^{-r^2} m \,d\theta\, dm\, n^5\, d\Omega_5 \,dn \\ \nonumber
&= \frac{(2\pi)(\pi^3)}{\mathcal{N}} \int  F\left(\frac{m}{r}\right) e^{-r^2} m\,n^5\, dm\, dn.
\end{align}
We have integrated out the angular coordinates $d\theta$ and $d\Omega_5$ over the 1- and 5-spheres respectively in the last line. If $F(|\psi_{ee}|)$ is linear in $|\psi_{ee}|^2$, as is the case for $F_{\text{finite }\Gamma}^{(1)}$, the final integral is easily performed, and the result is to replace $|\psi_{ee}|^2$ with $1/4$. Thus $F_{\text{finite }\Gamma}^{(1)}(\psi) = \bar{F}_{\text{finite }\Gamma}^{(1)}$ for any state with $|\psi_{ee}|^2=1/4$.

\begin{figure}
\centering
\includegraphics[width = \columnwidth]{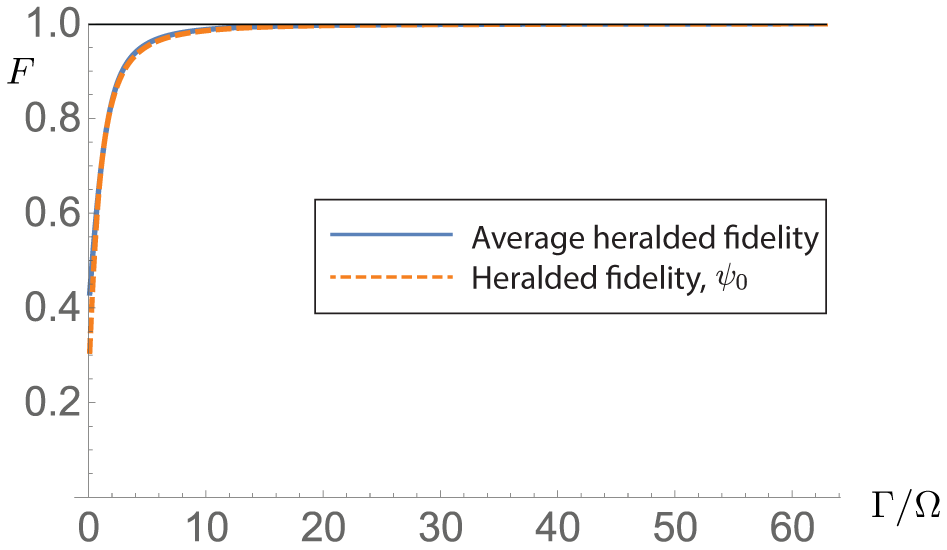}
\caption{Average heralded gate fidelity as computed numerically versus the heralded gate fidelity when applied to the specific state $|\psi_0\rangle$.}
\label{fig:FAverageHerald}
\end{figure}

The corresponding integral for $F_\text{herald}$ is non-trivial, so we evaluate it numerically. We plot the numerically computed $\bar{F}_\text{herald}$ against $F_\text{herald}(|\psi_0\rangle)$ in Fig.~\ref{fig:FAverageHerald}, with $|\psi_0\rangle = (|gg\rangle + |ge\rangle+|eg\rangle+|ee\rangle)/2$. We use the exact solution for $\psi_{ee}$, \erf{eq:PsiHerExact} for both computations. The curves overlap almost perfectly, indicating that the analytically tractable $F_\text{herald}(|\psi_0\rangle)$ serves as a good proxy for $\bar{F}_\text{herald}$. 

To derive the average gate fidelity $F_{\chi}$ due to finite $\chi/\kappa$, we use equation (17) of reference \cite{nielsen2002simple}, which computes the average fidelity between an ideal unitary operation $\hat{U}\,\hat{\rho}\,\hat{U}^\dagger$ and the output of some quantum channel $\mathcal{E}(\hat{\rho})$
\begin{align}
\bar{F}(\mathcal{E}, U) = \frac{\sum_i \Tr[\hat{U}\hat{U}_i^\dagger\hat{U}^\dagger \mathcal{E}(\hat{U}_i)] + d^2}{d^2(d+1)}
\end{align}
where $\hat{U}_i$ is matrix basis satisfying $\Tr[U_i U_j^\dagger] = \delta_{i,j} d$ and $d$ is the Hilbert space dimension. 
This is used to create Fig.~\ref{fig:2DAverageFid}, in conjunction with a steady--state approximation of the measurement described in Appendix \ref{app-largechi}. 
Specifically, we compute $\bar{F}_{\chi}$ (a measure of the fidelity that is retained despite RIP and decoherence effects \emph{within} the two--qubit subspace) by setting $\hat{U}=\hat{I}$ and using the two-qubit Pauli basis $\hat{\sigma}_k\otimes \hat{\sigma}_l$ with $\{k,l\}=0,x,y,z$. We apply \erf{eq:QubitMESolution} as $\mathcal{E}$. The result simplifies to \erf{eq:FbarFiniteChi}.

\section{Measurement Model Details \label{app-measurement}}

For the sake of completeness and pedagogy, we provide an overview of dispersive measurement, motivating the main quantities used to describe this measurement process the main text. 
The results and approaches outlined here are well--developed in the literature, and for further reading on the subject one may consult e.g.~\cite{blais2020circuit, gambetta2008trajectory, Blais2004, Korotkov2016, steinmetz2021, clerk2010noise, LeighShay2020}. 
We will here restrict discussion to measurement of a two--level system except where necessary.

\subsection{The Dispersive Hamiltonian}

Dispersive measurement is based on the Jaynes--Cummings Hamiltonian
\be \label{HJC}
\hat{H}_\mathrm{JC} = \tfrac{1}{2}\,\omega_q\,\hat{\sigma}_z + \omega_c \,\hat{a}^\dag \hat{a} + g\left(\hat{a}\, \hat{\sigma}^+ + \hat{a}^\dag\, \hat{\sigma}_- \right),
\ee
which is derived from the Rabi Hamiltonian (describing dipole coupling of a qubit to a quantized field mode) via the rotating wave approximation. 
We define the qubit--cavity detuning $\Delta^q_c \equiv \omega_q-\omega_c$, and move towards the dispersive regime defined by $\Delta^q_c \gg g$. 
One may then make a further approximation (for example, via second--order perturbation theory), to obtain the dispersive Hamiltonian
\be \label{Hdispersive}
\hat{H}_\mathrm{disp} = \frac{\omega_q}{2}\,\hat{\sigma}_z + \left( \omega_c + \frac{g^2}{\Delta^q_c}\,\hat{\sigma}_z \right)\hat{a}^\dag\hat{a}.  
\ee
We define the dispersive shift $\chi \equiv g^2/\Delta^q_c$, where in the notation of the main text, we have $\chi_e = \chi/2$ and $\chi_g = -\chi/2$.  
The dispersive interaction indicates that the field will experience an effective frequency shift depending on $\hat{\sigma}_z$ of the qubit; our aim here is to explain in detail how this can be leveraged to realize optical readout of the qubit's $\hat{\sigma}_z$ observable. 

\subsection{A Driven and Decaying Cavity}

Let us suppose that the cavity mode described by $\hat{a}$ above leaks into an external line $\hat{b}$ at a rate $\kappa$. 
That line will carry information to a detector in the ideal case, and we will say that information that has entered the line but is not detected has dissipated. 
It is important to appreciate that the cavity decay rate $\kappa$ functions as a gatekeeper, fundamentally limiting the rate at which information shared between the qubit and internal mode $\hat{a}$ becomes knowable from outside. 
Implicit in this description is the notion that our quantum system comprises both the qubit and internal cavity mode, which are completely isolated from the external world except for the coherent field input and decay into the line. 
This picture motivates us to use a Markovian description of the dissipation of the composite qubit $\otimes$ cavity system, even if the dynamics may manifest as non-Markovian when we try to describe the qubit by itself. 
To this end, we may picture the situation using the structure typical of a composite collision model \cite{ciccarello2021,Lorenzo2017}, illustrated in Fig.~\ref{fig:cQEDImplementation}c. 
At successive timesteps, the joint qubit--cavity system ``collides'' with a fresh temporal mode in an external line via cavity decay. 
Direct detection / monitoring of the line is assumed possible, but direct detection of the cavity is not, such that information becomes detectable only after decaying out of the cavity. 

We may model the cavity decay into the line over a time interval $\Delta t$ as a unitary beamsplitter relation
\be \label{decay-bs}
\hat{a} \quad\rightarrow\quad \sqrt{e^{-\kappa\,\Delta t}}\,\hat{a} + \sqrt{1-e^{-\kappa\,\Delta t}}\,\hat{b}_k,
\ee
where the operator $\hat{b}_k$ represents the line mode at the timestep indexed by $k$. The evolution of the internal cavity mode due to decay can equivalently be formulated as 
\be \label{io-decay}
\dot{a} = -\tfrac{1}{2}\kappa\,\hat{a}. 
\ee
Dispersive measurement requires that the cavity mode be supplied with some input amplitude as well. We may consequently supplement the equation \eqref{io-decay} (first) with an input $\varepsilon\,\hat{a}^\dag\,e^{i\,\omega_\varepsilon\,t+i\,\phi} + \varepsilon\,\hat{a}\,e^{-i\,\omega_\varepsilon\,t-i\phi}$, and (second) with the effect of the qubit on the cavity mode as per $\dot{a} = -i[\hat{a},\hat{H}_\mathrm{disp}]$.  
Combining these elements of the input/output picture yields \cite{Korotkov2016, clerk2010noise}
\be \label{io-full} 
\dot{a} = \varepsilon - \left(i\Delta_j + \tfrac{1}{2}\kappa\right) \hat{a},
\ee
where we have written the equation in the frame rotating at the input tone frequency $\omega_\varepsilon$, and we define $\Delta_j \equiv \omega_c-\omega_\varepsilon+\chi_j$ as in the main text. 
The block--diagonal structure of the dispersive Hamiltonian allows us to solve the equation in \emph{each} subspace $j$, and use them together as implied by Eq.~\eqref{eq:rhoPolaron}. 
We have again implicitly chosen a phase on $\varepsilon$ that corresponds to driving along the $\hat{P}$ quadrature ($\phi = \pi/2$, in the notation of the main text).

While many analyses stop here, we find that it is important in our present case to discuss Eq.~\eqref{outerphase} in greater detail.
This will be clearest if we consider a qubit--cavity state 
$\zeta\,e^{i\varphi_e}\ket{e,\alpha_e} + \upsilon\,e^{i\varphi_g}\ket{g,\alpha_g}$.
We may generally consider the $j^{th}$ subspace, where the element $e^{i\varphi_j}\ket{j,\alpha_j}$ of the joint state evolves under 
\be 
\hat{H}_j = \Delta_j\,\hat{a}^\dag\hat{a} + i\,\varepsilon\,\hat{a}^\dag - i\,\varepsilon\,\hat{a}.
\ee
We consider the short time evolution $\hat{U}_j(dt)\,e^{i\varphi_j} \ket{j,\alpha_j} = $
\begin{subequations} \label{eq-phasedot} \be 
e^{-i\,\hat{H}_j\,dt}\,e^{i\varphi_j} \ket{j,\alpha_j} \approx e^{i\varphi_j}\,e^{-i\,\Delta_j\,\hat{a}^\dag\hat{a}\,dt}\,\hat{D}(\varepsilon\,dt)\ket{j,\alpha_j},
\ee
where $\hat{D}(\beta) = e^{\beta\,\hat{a}^\dag - \beta^\ast\,\hat{a}}$ is the displacement operator, which we have separated from the other term (a phase rotation) only because higher--order terms in the Baker--Campbell--Hausdorff expansion scale to $O(dt^2)$ and higher.
Continuing, we may write $\hat{U}_j(dt)\,e^{i\varphi_j} \ket{j,\alpha_j}$
\be \begin{split}
& \approx e^{i\varphi_j+\varepsilon\,dt\,(\alpha_j^\ast - \alpha_j)/2}\,e^{-i\,\Delta_j\,\hat{a}^\dag\hat{a}\,dt}\,\hat{D}(\alpha_j + \varepsilon\,dt)\ket{j,0}
\\ & = e^{i\varphi_j-i\,\varepsilon\,dt(i\,\alpha_j^\ast-i\,\alpha_j)/2}\ket{j,(\alpha_j+\varepsilon\,dt)e^{-i\,\Delta_j\,dt}}
\\ & \approx e^{i\varphi_j - i\,\varepsilon\,dt\,\Im[\alpha_j]}\ket{j,\alpha_j + dt(\varepsilon - i\,\Delta_j\,\alpha_j)}.
\end{split} \ee \end{subequations}
The \emph{external} phase factor shows us that $\dot{\varphi}_j = -\varepsilon\,\Im[\alpha_j]$, while we may trivially recover \eqref{io-full} by looking at the coherent state evolution \emph{within} the ket, and then incorporating damping.
Assembling the calculations and arguments above, we have a time--continuum picture of the cavity dynamics under dispersive coupling, within each subspace defined by the qubit states.

\subsection{Qubit Decoherence and Measurement \label{app-qubitdisp}}

We next use the expressions above to write a family of Kraus--like operators which could be used to express the conditional (or un-conditional) evolution of the joint qubit--cavity state.
The approach below is most similar to those of Refs.~\cite{Lewalle_2021, FlorTeach, Korotkov2016, steinmetz2021}.
Following the intuition above, we write a qubit--cavity state of the form
\be \label{pure-form-state}
\ket{\psi_k} = \zeta_k\,e^{i\,\varphi_e^{(k)}} \ket{e,\alpha_e^{(k)}} + \upsilon_k\,e^{i\,\varphi_g^{(k)}} \ket{g,\alpha_g^{(k)}}; 
\ee
having already described the evolution of $\alpha$ and $\varphi$ above, our priority is now to understand the evolution of $\zeta$ and $\upsilon$.
Suppose that the initial qubit--cavity--line state reads $\ket{\psi_k}\otimes\ket{0}$ (recalling Fig.~\ref{fig:cQEDImplementation}),
and then evolves to 
\be \begin{split}
& \zeta_k\,e^{i\,\varphi_e} \ket{e, \sqrt{e^{-\kappa\,\Delta t}}\,\alpha_e^{(k)}} \otimes \ket{\sqrt{1-e^{-\kappa\,\Delta t}}\,\alpha_e^{(k)}} \\ & + \upsilon_k\,e^{i\,\varphi_g} \ket{g, \sqrt{e^{-\kappa\,\Delta t}}\,\alpha_g^{(k)}} \otimes \ket{\sqrt{1-e^{-\kappa\,\Delta t}}\, \alpha_g^{(k)}},
\end{split} \ee
as per \eqref{decay-bs} after decay into the line. 
We may introduce some idea of a final line state $\ket{L_k}$ corresponding to a possible detector outcome, and rearrange the above into a state update
\be
\underbrace{\bra{L_k} \left( \begin{array}{cc} 
\ket{\sqrt{1-e^{-\kappa\,\Delta t}}\,\alpha_e^{(k)}} & 0 \\
0 & \ket{\sqrt{1-e^{-\kappa\,\Delta t}}\,\alpha_g^{(k)}} 
\end{array} \right)}_{\hat{\mathcal{M}}} \ket{\psi_k}. 
\ee
We have now identified a Kraus--like operator $\hat{\mathcal{M}}$ which can be used to update the coefficients in \eqref{pure-form-state} \emph{given} that a detector at the end of the line obtains an outcome in timestep $k$ corresponding to $\ket{L_k}$.

The vector of $\zeta$, $\upsilon$ can be generalized to a density matrix $\rho$ (i.e.,~\eqref{pure-form-state} is more generally written as \eqref{eq:rhoPolaron}); this $\rho$ is expressible as a $2\times 2$ matrix due to the special form of the joint qubit--cavity state we have been able to assume. 
This means that as long as the only dynamics we consider are those due to measurement or dissipation via the dispersive channel described, the problem is not formally more complex than that of a qubit alone. 
Dispersive measurement is usually performed with homodyne (one quadrature) or heterodyne (two quadrature) detection at the end of the line. 
We do not need to consider the conditional evolution from either of these cases in detail however, and the interested reader should refer to e.g.~\cite{ steinmetz2021, LeighShay2020} for details. 
Here it will be adequate to consider the case where the line simply dissipates permanently, without detection, into the external environment. 
Any basis could be chosen to trace over all possible $\ket{L_k}$; we will use the Fock basis for simplicity. 
To $O(\Delta t)$, there are only two possible Fock--basis outcomes to a given timestep: Either zero photons emerge from the line ($\ket{L_k} = \ket{0}$), as per
\begin{subequations} \be \begin{split}
\hat{\mathcal{M}}_0 &= \left( \begin{array}{cc} 
 \ip{0}{\sqrt{\kappa\,\Delta t}\,\alpha_e}& 0 \\ 
0 &  \ip{0}{\sqrt{\kappa\,\Delta t}\,\alpha_g}
\end{array} \right) \\ &\approx \hat{\mathbb{I}}_2 + \Delta t \underbrace{\left( \begin{array}{cc} 
-\tfrac{1}{2}\kappa\,|\alpha_e|^2 & 0 \\ 0 & -\tfrac{1}{2}\kappa\,|\alpha_g|^2
\end{array}\right)}_{\hat{\mathfrak{Z}}}, \end{split} 
\ee 
or a single photon is detected ($\ket{L_k} = \ket{1}$), as per
\be \label{frakSdef} \begin{split}
\hat{\mathcal{M}}_1  &=  \left( \begin{array}{cc} 
\ip{1}{\sqrt{\kappa\,\Delta t}\,\alpha_e}  & 0 \\ 
0 & \ip{1}{\sqrt{\kappa\,\Delta t}\,\alpha_g} 
\end{array} \right) \\ & \approx \sqrt{\Delta t}\underbrace{
\left( \begin{array}{cc} \sqrt{\kappa}\,\alpha_e & 0 \\ 
0 & \sqrt{\kappa}\,\alpha_g \end{array} \right)}_{\hat{\mathfrak{S}}}.
\end{split} \ee \end{subequations}
Given $\rho(t) = \rho_k$, the unconditional state update to $\rho(t+\Delta t) = \rho_{k+1}$ may be written\footnote{Note that the effective non-Hermitian Hamiltonian \eqref{eq:HNoJump} describing the jumpless case may instead be derived from an expression like
$$ \rho_{k+1} = \frac{\hat{\mathcal{M}}_0\,\rho_k\,\hat{\mathcal{M}}_0^\dag}{\mathrm{tr}\left(\hat{\mathcal{M}}_0\,\rho_k\,\hat{\mathcal{M}}_0^\dag\right)}, $$
which is the state update conditioned on the outcome $0$. 
}
\be 
\rho_{k+1} = i\,\Delta t[\rho_k,\hat{H}_\varphi]+\hat{\mathcal{M}}_0\,\rho_k\,\hat{\mathcal{M}}_0^\dag + \hat{\mathcal{M}}_1\,\rho_k\,\hat{\mathcal{M}}_1^\dag + O(\Delta t^2),
\ee
or equivalently
\be \label{lindme}
\dot{\rho} = \frac{\rho_{k+1} - \rho_k}{\Delta t} = i[\rho,\hat{H}_\varphi] + \hat{\mathfrak{Z}}\,\rho + \rho\,\hat{\mathfrak{Z}}^\dag + \hat{\mathfrak{S}}\,\rho\,\hat{\mathfrak{S}}^\dag. 
\ee
We have included the resonator--induced phase (RIP, see \cite{Cross_2015, paik2016experimental}) with the diagonal Hamiltonian $\hat{H}_\varphi$ with matrix elements $H_{\varphi,jj} = \dot{\varphi}_j$.
It is, at this point, straightforward to recognize that $\hat{\mathfrak{S}}$ could be expressed in the form $\hat{\mathfrak{S}} = \hat{L}$, where $L_{jj} = \sqrt{\kappa}\,\alpha_j$ is a diagonal matrix with complex entries, and that $\hat{\mathfrak{Z}} = -\tfrac{1}{2}\hat{L}^\dag\hat{L}$ in that notation.
Therefore, \eqref{lindme} is just a Lindblad Master Equation (ME). 
This $\hat{L}$ generalizes the basic form $\sqrt{\Gamma}\,\hat{\sigma}_z$ implied by \eqref{eq:ZenoMasterEquation}, for the joint qubit--cavity state. 

By comparing \eqref{lindme} to the Lindblad ME for the $\hat{\sigma}_z$ channel of a qubit (i.e.,~$\hat{L} = \sqrt{\Gamma}\,\hat{\sigma}_z$), we may obtain a straightforward expression for the measurement strength (decoherence rate) as a function of $\alpha_e$ and $\alpha_g$. 
The dynamics \eqref{lindme} can be re-written in Bloch coordinates (again, generalizing $\zeta$ and $\upsilon$) via $\dot{q} = \mathrm{tr}(\hat{\sigma}_q\,\dot{\rho})$ for $q = x,y,z$, and read
\begin{subequations} \label{eq_app_analytic}
\be 
\dot{x} = -\Gamma\,x + \Upsilon\,y , \quad
\dot{y} = -\Gamma\,y - \Upsilon\,x , \quad
\dot{z} = 0,
\ee
where we have defined the measurement strength (dephasing rate)
\be \label{gammadef}
\Gamma \equiv \tfrac{1}{2}\kappa \left| \alpha_e - \alpha_g \right|^2 = \kappa\left(\tfrac{1}{2}|\alpha_e|^2 + \tfrac{1}{2}|\alpha_g|^2 - \Re[\alpha_e\,\alpha_g^\ast] \right),
\ee
and the average phase rotation rate (AC Stark shift and RIP)
\be \label{upsilondef}
\Upsilon = \kappa\,\Im[\alpha_e\,\alpha_g^\ast] + \varepsilon\,\Im[\alpha_g-\alpha_e]. 
\ee
\end{subequations}
These expressions have been derived in past investigations of the present problem \cite{Korotkov2016}, and clearly illustrate that a measure of the distance \eqref{gammadef} is the defining feature making information about $\ket{e}$ versus $\ket{g}$ available to the external world via the dispersive interaction. 
The expressions \eqref{meas-strength} and \eqref{ac-stark} in the main text are straightforward generalizations of the quantities we have just derived. 
The effective damping and rotation rates of the density matrix may be further modified if one traces out the cavity as well as the line \cite{Korotkov2016} in the NM case.

\subsection{Two Dispersive Markovian Regimes \label{app-Markov}}

\subsubsection{The Fast Markovian Regime} 

The Markovian regime is typically defined by $\kappa$ being the fastest timescale directly involved in the measurement process. 
Suppose we measure ``continuously'' in such a way as to obtain a measurement outcome every $\Delta t$, due to some detector integration over that same interval. 
We may more precisely point to a hierarchy of timescales associated with this Markovian regime, that make it ideal for measurement: First, we suppose that $\Delta t \gtrsim \kappa^{-1}$ (so that the cavity has time to approximately reach its steady state within every measurement interval). 
If we add unitary rotations, we must secondly suppose that $\kappa \gg \Omega$ (so that the cavity is able to quasi-adiabatically follow qubit dynamics on timescales of interest). Implicit above is the notion that $\Delta t\,\Omega \ll 1$.

The measurement strength \eqref{gammadef} reads  
\be \begin{split}
\bar{\Gamma} &= \frac{8\,\kappa\,\varepsilon^2\,(\chi_e-\chi_g)^2}{(\kappa^2+4 \Delta_e^2)(\kappa^2 + 4 \Delta_g^2)} = \frac{32\,\kappa\,\varepsilon^2\,\chi^2}{(\kappa^2+4 \Delta_e^2)(\kappa^2 + 4 \Delta_g^2)},
\end{split} \ee
if we assume the cavity has reached its steady state \eqref{cavss} within the measurement interval. 
In the limit $\kappa \gg \Delta_j$, it is possible to rewrite this as
\be \label{GammaBar}
\bar{\Gamma} \approx \frac{32\,\varepsilon^2\,\chi^2}{\kappa^3} \approx \frac{8\,\chi^2}{\kappa}\langle n \rangle,
\ee
where the last form is in agreement with the initial derivations of dispersive readout by Blais et al.~\cite{Blais2004}, and where $\langle n \rangle$ is the average photon number in the cavity. 
This last expression \eqref{GammaBar} is typically used in the Markovian regime. 
It offers clear intuition about the essential dependence of the measurement strength on the dispersive shift, cavity decay rate, and photon number populating the measurement tone, in the regime best suited to straightforward continuous measurement of the qubit \cite{gambetta2008trajectory, Murch:2013ur}. 

\subsubsection{The Slow Markovian Regime}

The usual fast Markovian regime above is but one steady state approximation we can make however. 
For the Zeno gate, it is helpful to also investigate a \emph{slow} limit, defined instead by the hierarchy $\Omega \ll \kappa \ll \Delta \chi$. 
We call this a slow limit, because it is difficult in practice to make the relative dispersive shifts $\Delta\chi$ arbitrarily large; consequently, $\kappa$ is limited, and any unitary dynamics at rate $\Omega$ must in turn be slower, such that the hierarchy of timescales above implies a very long gate time $T_G  = 2\pi/\Omega$.

Let us briefly look at this limit in the context of a qutrit, where we will have some $\chi_f$, $\chi_e$, and $\chi_g$. 
The steady state measurement strength reads 
\be \label{eq-Gammaij-steady}
\bar{\Gamma}_{ij} = \frac{8\,\kappa\,\varepsilon^2\,\Delta\chi_{ij}^2}{(\kappa^2 + 4\Delta_i^2)(\kappa^2 + 4\Delta_j^2)}
\ee
for $\Delta_j = \Delta^c_\varepsilon + \chi_j$ and $\Delta\chi_{ij} \equiv \chi_i - \chi_j$.  
Now suppose that in analogy with the main text, we choose our Zeno tone frequency such that $\Delta_f = 0$, with the intent that $\bar{\Gamma}_{fe} \sim \bar{\Gamma}_{fg} \gg \bar{\Gamma}_{eg}$. 
It follows from that choice of $\omega_\varepsilon$ that $\Delta_e \rightarrow \Delta\chi_{fe}$ and $\Delta_g \rightarrow \Delta\chi_{fg}$. 
Putting these results into the expressions for the steady--state measurement rate for each of three qutrit transitions, and approximating for $\kappa \ll \Delta \chi$, we find
\begin{subequations} \label{eqs-GammaSS-qutrit}
\be 
\bar{\Gamma}_{eg} = \frac{8\,\kappa\,\varepsilon^2\,\Delta\chi_{eg}^2}{(\kappa^2 + 4\Delta\chi_{fe}^2)(\kappa^2 + 4\Delta\chi_{fg}^2)} \approx \frac{\kappa\,\varepsilon^2}{2} \frac{\Delta \chi_{eg}^2}{\Delta \chi_{fe}^2 \Delta \chi_{fg}^2}
\ee 
for the case where $\omega_\varepsilon$ is not resonant with either state's dispersive shift, and
\be \label{eq-GAM-strong-markov}
\begin{split}
\bar{\Gamma}_{fe} = \frac{8\,\kappa\,\varepsilon^2\,\Delta\chi_{fe}^2}{\kappa^2(\kappa^2 + 4\Delta\chi_{fe}^2)} &\approx \frac{2\,\varepsilon^2}{\kappa} \\& \approx  \frac{8\,\kappa\,\varepsilon^2\,\Delta\chi_{fg}^2}{\kappa^2(\kappa^2 + 4\Delta\chi_{fg}^2)} = \bar{\Gamma}_{fg} 
\end{split} \ee \end{subequations}
for transitions to $\ket{f}$, which differ due to the resonance $\Delta_c^\varepsilon = \chi_f$. 
These expressions correspond to the regime of a Markovian but very slow Zeno gate. 
The especially desirable property of the expressions above is that
\be \label{slow_approx_order}
\bar{\Gamma}_\mathrm{Zeno} \sim O\left(\frac{\varepsilon^2}{\kappa} \right) \gg \bar{\Gamma}_{\mathcal{S}_Z}\sim O\left(\frac{\kappa\,\varepsilon^2}{\Delta \chi^2}\right) 
\ee
(where the Zeno subscript refers to transitions we wish to block, and $\mathcal{S}_Z$ refers to the computational subspace which we wish to leave alone). 
The validity of these approximations is based on $T_G\,\kappa/2\pi \gg 1$ (where $T_G = 2\pi/\Omega$ is the gate time). 
Note that the restrictions on cavity photon number necessary to the dispersive regime prevents us from making $\varepsilon$ arbitrarily large. 
Generalization of these expressions to larger systems is quite simple. 

Using the same approach as above, we may also write approximate expressions for the AC--Stark and RIP rotations to leading order, obtaining
\begin{subequations} \label{eqs-UpsilonSS-qutrit} \be  
\bar{\Upsilon}_{eg} \approx - \frac{\varepsilon^2 \, \Delta \chi_{eg}}{\Delta \chi_{fg} \Delta \chi_{fe}},
\ee \be 
\bar{\Upsilon}_{fe} \approx - \frac{3\,\varepsilon^2}{\Delta \chi_{fe}}, \quad \bar{\Upsilon}_{fg} \approx - \frac{3\,\varepsilon^2}{\Delta \chi_{fg}}. 
\ee \end{subequations}

\subsection{Towards the Zeno Gate: Analytic Treatment of Qutrit Measurement \label{app-qutrit}}

We first introduce some notations and conventions for our discussion of qutrit states and measurement, and then proceed move towards the $\Delta \chi \gg \kappa$ limit. 

We choose to parameterize the state of a qutrit $\otimes$ cavity via the generalized Gell--Mann matrices
\begin{widetext} \begin{subequations} \be 
\hat{\sigma}_1 = \ket{f}\bra{f} - \ket{e}\bra{e} =  \left( \begin{array}{ccc} 1 & 0 & 0 \\ 0 & -1 & 0 \\ 0 & 0 & 0 \end{array} \right), \quad \hat{\sigma}_2 = \frac{1}{\sqrt{3}}\left( \begin{array}{ccc} 1 & 0 & 0 \\ 0 & 1 & 0 \\ 0 & 0 & -2 \end{array} \right), 
\ee \be 
\hat{\sigma}_3 = \left( \begin{array}{ccc} 0 & 1 & 0 \\ 1 & 0 & 0 \\ 0 & 0 & 0 \end{array} \right), \quad 
\hat{\sigma}_4 = \left( \begin{array}{ccc} 0 & 0 & 1 \\ 0 & 0 & 0 \\ 1 & 0 & 0 \end{array} \right), \quad 
\hat{\sigma}_5 = \left( \begin{array}{ccc} 0 & 0 & 0 \\ 0 & 0 & 1 \\ 0 & 1 & 0 \end{array} \right),
\ee \be 
\hat{\sigma}_6 = \left( \begin{array}{ccc} 0 & -i & 0 \\ i & 0 & 0 \\ 0 & 0 & 0 \end{array} \right), \quad 
\hat{\sigma}_7 = \left( \begin{array}{ccc} 0 & 0 & -i \\ 0 & 0 & 0 \\ i & 0 & 0 \end{array} \right), \quad 
\hat{\sigma}_8 = \left( \begin{array}{ccc} 0 & 0 & 0 \\ 0 & 0 & -i \\ 0 & i & 0 \end{array} \right).
\ee \end{subequations}
It follows that the qutrit density matrix can be expressed via these matrices and the corresponding coordinates, as per
\be 
\rho = \tfrac{1}{3} \,\hat{\mathbb{I}}_3 + \tfrac{1}{2}\, \mathbf{q}\cdot\hat{\boldsymbol{\sigma}},
\ee
where we have defined the Bloch--like coordinates $\mathbf{q} = \mathrm{tr}(\hat{\sigma}_\mathbf{q}\,\rho)$. 
If we na\"{i}vely supplement the dynamics given by the Lindblad ME with Rabi rotations between the $\ket{e}$ and $\ket{f}$ levels, as per $\hat{H}_R = \tfrac{1}{2}\,\Omega\,\hat{\sigma}_6$, then we obtain the dynamical equation
\begin{subequations} \label{qutrit-me}
\be \begin{split}
\dot{\rho} = i[\rho,\hat{H}_R+\hat{H}_\varphi] + &\hat{\mathfrak{Z}}\,\rho + \rho\,\hat{\mathfrak{Z}}^\dag + \hat{\mathfrak{S}}\,\rho\,\hat{\mathfrak{S}}^\dag \\ \text{for} \quad \hat{\mathfrak{Z}} = - \frac{\kappa}{2} \left( \begin{array}{ccc}
|\alpha_f|^2 & 0 & 0 \\
0 & |\alpha_e|^2 & 0 \\
0 & 0 & |\alpha_g|^2 
\end{array} \right) \quad&\text{and}\quad \hat{\mathfrak{S}} = \sqrt{\kappa}\left( \begin{array}{ccc}
\alpha_f & 0 & 0 \\
0 & \alpha_e & 0 \\
0 & 0 & \alpha_g
\end{array} \right),
\end{split} \ee
which can equivalently be written in terms of the coordinates just defined:
\be 
\dot{q}_1 = -\Omega\,q_3,\quad \dot{q}_2 = 0,
\ee \be 
\dot{q}_3 = \Omega\,q_1 - \Gamma_{fe}\,q_3 + \Upsilon_{fe}\,q_6,\quad 
\dot{q}_4 = -\tfrac{1}{2}\Omega\,q_5 - \Gamma_{fg}\,q_4 + \Upsilon_{fg}\,q_7,\quad 
\dot{q}_5 = \tfrac{1}{2}\Omega\,q_4 - \Gamma_{eg}\,q_5 + \Upsilon_{eg}\,q_8,
\ee \be 
\dot{q}_6 = -\Gamma_{fe}\,q_6 - \Upsilon_{fe}\,q_3,\quad 
\dot{q}_7 = -\tfrac{1}{2}\Omega\,q_8 - \Gamma_{fg}\,q_7 - \Upsilon_{fg}\,q_4,\quad 
\dot{q}_8 = \tfrac{1}{2}\Omega\,q_7 - \Gamma_{eg}\,q_8 - \Upsilon_{eg}\,q_5.
\ee\end{subequations}
\end{widetext}
Note that we have again collected the resonator--induced phase rotations $\dot{\varphi}_j = - \varepsilon\,\Im[\alpha_j]$ into a Hamiltonian term $\hat{H}_\varphi = -\varepsilon \sum_j \Im[\alpha_j]\,\hat{\Pi}_j$ (this is essentially a phase gate in the joint qutrit--cavity space). 
We remind the reader that the expressions immediately above are written under the assumption that $\mathbf{q}$ parameterizes a \emph{joint} qutrit--cavity state of the form \eqref{eq:rhoPolaron}, such that the line is traced out but the cavity is not.  
Then $\Gamma$ and $\Upsilon$ have the same definitions as in \eqref{gammadef} and \eqref{upsilondef}, with the indices denoting the pair of levels associated with the coefficient. 
We reiterate that any allusions to a na\"{i}ve application of the CA, as used in e.g.~Figs.~\ref{fig-ideal-chi}--\ref{fig-res-chi-slow}(a,b), refer the same process leading to \eqref{qutrit-me}, and that the na\"{i}ve part comes from appending $i[\rho,\hat{H}_R]$ to the equation of motion for measurement only \emph{without} further accounting for how those unitary dynamics and measurement process impact each other. 
Notice that while the Rabi drive directly causes rotations $q_1 \leftrightarrow q_3$ at frequency $\Omega$, $\Omega$ also indirectly generates dynamics elsewhere; analogs of the latter types of terms are important in the Zeno gate of the main text operating in a larger space.
The rotations $\Upsilon$, on the other hand, constitute an extra complication that is particular to the physical implementation under consideration.

We may look at the qutrit equations of motion explicitly in the $\Omega \ll \kappa \ll \Delta \chi$ regime. 
Recall that we computed approximate expressions for $\Gamma$ and $\Upsilon$ in  \eqref{eqs-GammaSS-qutrit} and \eqref{eqs-UpsilonSS-qutrit}, respectively. 
Combining these with our previous expressions, we may understand the main aspects of the dynamics by focusing on the evolution of the real parts of the coherences:
\begin{subequations} \be \label{q3dot_approx}
\dot{q}_3 \approx \Omega\,q_1 - \varepsilon^2 \left(\frac{2\,q_3}{\kappa} + \frac{3\,q_6}{\Delta \chi_{fe}} \right),
\ee \be \label{q4dot_approx}
\dot{q}_4 \approx -\tfrac{1}{2}\Omega\,q_5 - \varepsilon^2\left( \frac{2\,q_4}{\kappa} + \frac{3\,q_7}{\Delta \chi_{fg}} \right),
\ee \be \label{q5dot_approx}
\dot{q}_5 \approx \tfrac{1}{2}\Omega\,q_4 - \frac{\varepsilon^2\,\Delta\chi_{eg}}{\Delta \chi_{fe} \Delta \chi_{fg}}\left(\frac{\kappa\,\Delta \chi_{eg}}{2\,\Delta \chi_{fe} \Delta \chi_{fg}}\,q_5  + q_8\right) . 
\ee \end{subequations}
While \eqref{slow_approx_order} ensures that coherences connecting to $\ket{f}$ ($q_3$ and $q_4$) damp much faster than those which do not (i.e.,~$q_5$), there is one moderately inconvenient feature apparent in these dynamics: 
While \eqref{q3dot_approx} and \eqref{q4dot_approx} exhibit faster damping than AC--Stark / RIP rotations ($\kappa^{-1} \gg \Delta \chi^{-1}$, which is good), 
\eqref{q5dot_approx} illustrates that the AC--Stark / RIP effects are not mitigated within the computational subspace to the same degree that the damping is. Practical implementations of the Zeno gate may consequently require either extra controls to mitigate phase rotations (see Ref.~\cite{blumenthal2021} for an example of this), or may account for RIP dynamics and use them alongside the Zeno dynamics emphasized here to achieve useful non-local operations.
For the Zeno gate, the best dynamics we could wish for above are those which are recovered in the limit $\Delta \chi_{eg} \rightarrow 0$ (i.e.,~$\Delta \chi_{\mathcal{S}_Z} \rightarrow 0$), which we have used as an idealized reference point in Fig.~\ref{fig-ideal-chi}. 
We finally remark that $\varepsilon^2$ scales against all the terms of interest above in the same way, and consequently has a double--edged effect: Both desired and undesired damping, and AC--Stark rotations, are amplified with larger $\varepsilon$, so that both our Zeno blocking and sources of gate infidelity grow together with increased tone power.

\subsection{Analysis: Zeno gate with large dispersive shifts \label{app-largechi}} 

We continue by using the approximations we have just introduced to make analytic statements about the Zeno gate fidelity in the $\Delta\chi \gg \kappa$ limit.  
Generalizing to a qutrit $\otimes$ $N$--qubit system, we may note that under the cavity steady--state assumptions detailed above, we have qudit evolution 
\begin{subequations}\be 
\dot{\rho}_{j\ell} = \tfrac{1}{2}(i\,\bar{\Upsilon}_{j\ell} - \bar{\Gamma}_{j\ell})\rho_{j\ell} \quad\text{for}\quad j\neq\ell, \quad \dot{\rho}_{jj} = 0.
\ee 
Consequently, all of the evolution is damping of coherences (dephasing), and rotations (due to $\hat{H}_R$ and the AC Stark effect), such that 
\be  \label{eq:QubitMESolution}
\rho_{j\ell}(t) = \rho_{j\ell}(0)\,e^{(i\,\bar{\Upsilon}_{j\ell}\,t - \bar{\Gamma}_{j\ell}\,t)/2}
\ee \end{subequations}
(where $\Upsilon_{jj}$ and $\Gamma_{jj}$ are zero by definition). 
We may use solutions of this form to compute the heralded and un-heralded Zeno gate fidelities in the slow Markovian regime, used to create Fig.~\ref{fig:2DAverageFid}. 
We reiterate that the above neglects the action of $\hat{H}_R$, which does not commute with the action of \erf{eq:QubitMESolution}, but that the ad-hoc addition of $\hat{H}_R$ is the least problematic in the Markovian regime we explicitly consider here \cite{koolstra2021monitoring}.
We add the unitary rotations at frequency $\Omega$ in this way, and then compare the approximate dynamics above against the ideal dynamics $\hat{U}_\mathrm{Zeno}$ \eqref{eq:DeriveUZeno} to compute the fidelity. 

To relate the subsequent expression to the results of the previous section, we substitute $t=2\pi/\Omega$ and $\varepsilon^2 = \Gamma \kappa/2$ [recall \eqref{eq-GAM-strong-markov}]. 
The resulting expressions now only depend on the unitless ratios $\Delta\chi/\kappa \equiv X$ and $\Gamma/\Omega \equiv Y$, which are the relevant physical quantities for the gate fidelity. Computing the average gate fidelity \cite{nielsen2002simple} in Appendix \ref{sec:Haar} (due to decoherence and phase shifts within the two--qubit computational subspace $\mathcal{S}_Z$), we find 
\be \begin{split} \label{eq:FbarFiniteChi}
\bar{F}_\chi = \frac{1}{2} &+ \frac{1}{10} \,\exp\left[ -\frac{\pi\,Y}{9\,X^2} \right] \cos\left[ \frac{\pi \,Y}{3\,X} \right] \\ & + \frac{1}{5} \,\exp\left[ -\frac{\pi\,Y}{16\,X^2} \right] \cos\left[ \frac{\pi \,Y}{4\,X} \right] \\ & + \frac{1}{5} \,\exp\left[ -\frac{\pi\,Y}{144\,X^2} \right] \cos\left[ \frac{\pi \,Y}{12\,X} \right].
\end{split} \ee
Examining the above expression in isolation, one would conclude they ought to maximize $\chi/\kappa$ and minimize $\Gamma/\Omega$. 
However from Sec. \ref{sec:FiniteGamma}, we know that it is also preferable to maximize $\Gamma/\Omega$ to suppress leakage through the Zeno block, implying a trade-off between a large measurement rate and spurious dephasing and Stark shifts. In Fig.~\ref{fig:2DAverageFid}, we plot the infidelity due to \erf{eq:FbarFiniteChi}, 
with the additional subtraction of the infidelities calculated in Sec. \ref{sec:FiniteGamma}. 
Fig.~\ref{fig:2DAverageFid} illustrates these ideas, where we plot
\begin{subequations} \label{eq-FID-fig5} \be 
\bar{F}_\chi - (1 - {F}_\mathrm{finite~\Gamma}) - (1 - {F}_\mathrm{herald}) \approx \bar{F}_\chi\cdot {F}_\mathrm{finite~\Gamma}\cdot {F}_\mathrm{herald}
\ee 
for the un-heralded case, and
\be 
\bar{F}_\chi - (1 - {F}_\mathrm{herald}) \approx \bar{F}_\chi\cdot{F}_\mathrm{herald}
\ee\end{subequations}
for the heralded case. This figure allows to 1) ascertain the degree of improvement in the heralded case as compared with the un-heralded one, and 2) quickly locate the parameter choices with the best trade-off between the different sources of infidelity in the expressions above. 
This also 3) unifies the different fidelity expressions derived at different points in the text to the extent possible.

\subsection{Comparison to Longitudinal Readout} 

Suppose that in place of the approximate \eqref{HJC}, derived from ``transverse'' dipole coupling between qubit and field, we instead had system with a ``longitudinal'' coupling \cite{blais2020circuit}
\be \label{H_longitude_qubit}
\hat{H}_L = \tfrac{1}{2}\,\omega_q\,\hat{\sigma}_z+\omega_c\,\hat{a}^\dag\hat{a} + g_z(\hat{a}+\hat{a}^\dag)\hat{\sigma}_z. 
\ee
The ansatz \eqref{pure-form-state} remains appropriate to describe measurement based on this qubit--field interaction alone, and the cavity evolution is now characterized by
\begin{subequations}\be 
\dot{a} = -i[\hat{a},\hat{H}_L] - \tfrac{1}{2}\,\kappa\,\hat{a} = g_z\,\hat{\sigma}_z -\left(i\,\omega_c + \tfrac{1}{2}\,\kappa   \right) \hat{a},
\ee
or equivalently
\be
\dot{\alpha}_\pm = -\left(i\,\omega_c + \tfrac{1}{2}\,\kappa   \right) \alpha_\pm \mp i\,g_z
\ee with solutions
\be \label{alpha-longit}
\alpha_\pm(t) = \left(\alpha_\pm(0)\mp\frac{2i\,g_z}{\kappa+2i\,\omega_c}\right)e^{-\left(i\,\omega_c+\tfrac{1}{2}\kappa\right)t}\mp\frac{2i\,g_z}{\kappa+2i\,\omega_c},
\ee \end{subequations} 
where $+$ is for $\ket{e}$ and $-$ is for $\ket{g}$. 

A major difference between the dispersive and longitudinal solutions is that the latter \emph{does not} imply the same tradeoffs for timescales in the measurement dynamics. 
Let us reconsider Markovian qutrit measurement for longitudinal coupling, following the same process employed in Eqs.~\eqref{eq-Gammaij-steady}--\eqref{eqs-GammaSS-qutrit}. 
Notice that $\kappa$ still sets the rate at which the cavity approaches a steady state, and that $\Omega \ll \kappa$ should still be a suitable criterion for approximately Markovian and coherent state dynamics. 
However, to make approximations similar to those in Eqs.~\eqref{eq-Gammaij-steady}--\eqref{eqs-GammaSS-qutrit}, we now require the hierarchy $\Omega \ll \kappa \ll \omega_c$ instead of $\Omega \ll \kappa \ll \Delta \chi$, where $\omega_c$ is the bare cavity frequency. 
As $\omega_c$ can be on the order of $\mathrm{GHz}$ in experiments, we find that our restrictions on $\kappa$ and the gate time are much more relaxed for longitudinal coupling compared with the dispersive case. 
The steady state measurement strength here reads
\be \label{eq-Gammaij-longit}
\bar{\Gamma}_{j\ell} = \frac{2\,\kappa\,\Delta g_{j\ell}^2}{\kappa^2 + 4\,\omega_c^2} \approx \frac{\kappa\,\Delta g_{j\ell}^2}{2\,\omega_c^2},
\ee
where the latter approximation is based on $\kappa \ll \omega_c$. Notice that in practice, we do \emph{not} want to go deep into this regime if we want a strong measurement rate; the expressions above suggest that for a strong Zeno gate measurement we want a very quickly--decaying cavity, as modest a cavity resonance frequency as possible, and the largest $\Delta g_{j\ell}$ we can manage.
The relative measurement strengths between different transitions $\ket{j} \leftrightarrow\ket{\ell}$ are now entirely determined by the differences between coupling terms $\Delta g_{j\ell}^2$. 
With the longitudinal approach, one's ability to engineer suitable $g_j$ is thus the limiting factor for both the Zeno blocking capability and subspace decoherence. 
If those $g_j$ can be engineered precisely and freely, then the longitudinal scheme leaves considerable latitude to choose a gate time that optimizes tradeoffs between Zeno blocking and subspace decoherence.

\begin{figure*}
\includegraphics[width = \textwidth]{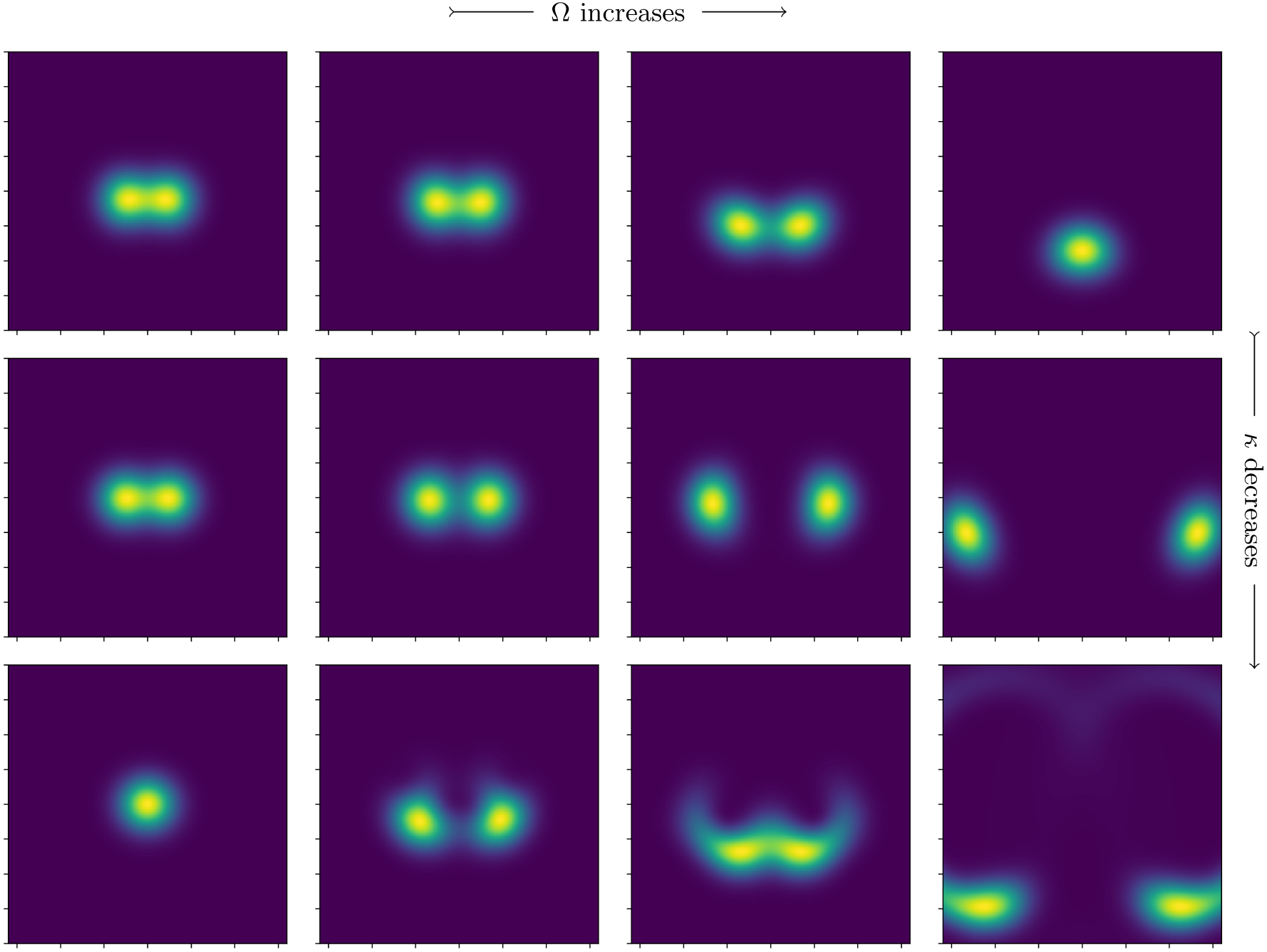}
\caption{
We plot the Husimi--$\mathcal{Q}$ \eqref{Husimi} functions of the readout resonator (with $X \in [-8,8]$ on $x$--axes and $P \in [-8,8]$ on the $y$-axes for all panels).  
We use $\varepsilon/2\pi = 2\,\mathrm{MHz}$, with dispersive shifts $\chi/2\pi = \pm 2\,\mathrm{MHz}$ (with $+$ for $\ket{e}$ and $-$ for $\ket{g}$), initialize the qubit in $(\ket{e}+\ket{g})/\sqrt{2}$, and trace out the qubit state to generate a plot after $T = 5\,\mathrm{\mu s}$. 
The readout tone of amplitude $\varepsilon$ is driven at the natural cavity frequency (i.e.,~the dispersive shifts are distributed about it symmetrically) for simplicity.  
The Rabi drive $\Omega$ is varied as we move from left to right as $\Omega = 0$ (far left column), $\Omega/2\pi = 5\,\mathrm{MHz}$ (center left column), $\Omega/2\pi = 50\,\mathrm{MHz}$ (center right column), and $\Omega/2\pi = 500\,\mathrm{MHz}$ (far right column). 
Similarly, we scale the cavity decay rate from top to bottom as $\kappa = 10\,\mathrm{MHz}$ (top row), $\kappa = 1\,\mathrm{MHz}$ (center row), and $\kappa = 0.1\,\mathrm{MHz}$ (bottom row). 
Essentially this puts the ``most Markovian'' dynamics in the top left, and ``least Markovian'' dynamics in the bottom right. 
The left column shows the case where the analytic measurement model \eqref{eq_app_analytic} is exact, and we essentially scale from this towards the decoupled dynamics \cite{Szombati2020} as we move right; this no--measurement limit (defined by $\Omega \gtrsim \Delta \chi^2/\kappa$) is attained for the top right panel only. 
Distortion of the initially--coherent cavity pointer states becomes increasingly clear as the cavity slows (i.e.,~towards the bottom right).  
Reading down the far left column, we see the weakening of the measurement due to increasingly slow cavity ring--up (recall Fig.~\ref{fig-linestrength}(c,d)). 
The most important takeaway from these figures, with respect to interpreting the dynamics of the main text, is that in any given row, turning up the Rabi rotation rate (i.e.,~scanning left to right) leads to \emph{an increase} in the pointer state distance, and hence a \emph{strengthening} of a measurement compared to the analytic model, before the rotations get too fast and qubit information is effectively wiped from the cavity. Such effects are wholly absent from the analytic measurement model. 
This apparent effect may explain the discrepancy between the analytic model (i.e.,~using the measurement strength as computed based dynamics shown in the left column) versus simulation, apparent in Figs.~\ref{fig-ideal-chi}--\ref{fig-res-chi-slow}. 
For example, the decrease in measurement strength (pointer state distance) in reading down the far left column is notably \emph{not} apparent reading down the center left column. 
This indicates that the introduction of $\Omega$, which acts only on the qubit, can indirectly lead to the appearance of a new timescale in the cavity evolution --- an effect which is completely absent from na\"{i}ve use of the CA model with non-negligible $\Omega$.
} \label{fig-Qfunc_2xCav}
\end{figure*}

Another significant difference between the longitudinal and dispersive solutions is that it possible to have $\Upsilon = 0$. 
This is a benefit in many measurement contexts (see \cite{blais2020circuit} and references therein), including ours: The phase rotations from $\Upsilon \neq 0$ interfere with our ideal Zeno gate dynamics, and consequently suppressing them would simplify practical implementation of the scheme considerably. 
Recall that there are two terms which may contribute to $\Upsilon$ (in direct analogy with \eqref{eq-phasedot}). 
If all the $g_j$ are real, then coherent states are always displaced along a line in the quadrature plane; without relative rotations we have $\kappa\,\Im[\bar{\alpha}_j\bar{\alpha}_\ell^\ast] = 0$, thereby eliminating one contribution to $\Upsilon_{j\ell}$.
In analogy with the RIP gate \cite{Cross_2015, paik2016experimental}, we still have a second relative phase contribution; considering again superpositions of terms $e^{i\varphi_j}\ket{j,\alpha_j}$, we find phase accumulation between different subspaces at rate $\Upsilon_{j\ell} = \dot{\varphi}_j - \dot{\varphi}_\ell = g_\ell\,\Re[\alpha_\ell] - g_j\,\Re[\alpha_j]$ (assuming real $g$). This may be obtained via the same logic as in Eq.~\eqref{eq-phasedot}.
With relaxed contraints on $\kappa$, we may again look at the steady--state solutions in the fast Markovian regime, and find
\be 
\bar{\Upsilon}_{j\ell} = \frac{4\omega_c (g_j^2-g_\ell^2)}{\kappa^2 + 4\omega_c^2} \approx \frac{g_j^2-g_\ell^2}{\omega_c}. 
\ee
This implies that a fast gate is possible with longitudinal readout, but that phase rotations are likely to remain a problem unless a specific situation can be engineered: We may have $\bar{\Gamma}_{j\ell} \gg 0$ and $\bar{\Upsilon}_{j\ell} = 0$ \emph{if} $g_j = -g_\ell$. 
Fig.~\ref{fig-ideal-lon} is based on this assumption, demonstrating ideal Zeno gate performance for large $g_{fe} = |g|$, $g_j = -|g|\,\forall\,j \neq fe$, and suitable choices of $\kappa$.


\section{Numerical Investigations Beyond the Markovian Regime \label{app-moresim}}

We here supplement the arguments of the main text by generating a few additional figures characterizing aspects of the dispersive measurement dynamics pertinent to the Zeno gate. 
As in the main text, numerics are based on the dispersive Hamiltonian \eqref{eq_HDisp} and Lindbladian dissipation of the cavity into an external line or waveguide \eqref{eq:OpenCavME}, the latter of which could be monitored. We implement these models using QuTip's \texttt{mesolve} function.

We here explicitly consider the \emph{cavity} dynamics which are an intrinsic part of the dispersive measurement process. We will be able to see the main features of interest focusing on a two--level system.
We use the Husimi--$\mathcal{Q}$ functions
\be \label{Husimi}
\mathcal{Q}(\alpha) = \tfrac{1}{\pi} \bra{\alpha} \mathrm{tr}_\mathrm{qubit}(\hat{\rho}) \ket{\alpha}\, \quad\text{with}\quad \alpha = X + iP
\ee
to represent the cavity state, with the qubit degree of freedom traced out.

Fig.~\ref{fig-Qfunc_2xCav} confirms that we can expect the coherent--state structure of the analytic measurement model to be only moderately distorted in the shallow NM regime. 
Note that $\Omega = 0$ is used in the left column of Fig.~\ref{fig-Qfunc_2xCav}, reproducing the analytic measurement model exactly. 
Several effects not accounted for in the analytic model are apparent is we introduce $\Omega$ and move into the NM regime, however: 
In particular, we observe (first) effective \emph{increases} in the distance between phase space features (which correlates with an effective measurement rate) with growing Rabi drive $\Omega$, before (second) $\Omega$ gets so large that the qubit and cavity essentially become decoupled \cite{Szombati2020}. 
Third, distortions of the coherent state pointers assumed in our analytic measurement models become apparent, especially for slower $\kappa$. 

We highlight these features in specific panels of Fig.~\ref{fig-Qfunc_2xCav}. 
Decoupling (``quantum rifling'' \cite{Szombati2020}) appears in the top right panel. 
Drive--induced \emph{enhancement} of the cavity dynamics are moderately apparent for smaller $\Omega$ in the top row however, and are extremely apparent in the middle row: Even when $\kappa$ is too slow to allow substantial cavity dynamics over a given time interval, a larger $\Omega$ can actually generate some cavity dynamics on a faster timescale. 
This potentially explains why our simulations in the main text do not show a steep drop in measurement strength for very slow $\kappa$, despite the very slow measurement ring--up time predicted by the CA model (recall the behavior in Fig.~\ref{fig-linestrength}). 
This effect is apparent in third row as well, where it occurs in conjunction with more exaggerated distortion of the initially coherent states.

\bibliographystyle{quantum}
\bibliography{Bibliography}
\end{document}